\documentclass[journal,9pt]{IEEEtran}
\usepackage{changes}
\usepackage{cite}
\usepackage{amsmath,amssymb,amsfonts, amsthm}
\usepackage{algorithmic}
\usepackage{graphicx}
\usepackage{textcomp}
\usepackage{colortbl}
\definecolor{shadegray}{RGB}{211,211,211}
\definecolor{lightgray}{RGB}{130,130,130}
\usepackage{float}
\usepackage{color}
\usepackage{xcolor}
\usepackage[colorlinks,linkcolor=red,anchorcolor=blue,citecolor=green]{hyperref}
\usepackage{nicefrac}
\usepackage{multirow}
\usepackage{cleveref}
\usepackage{makecell}
\usepackage{booktabs}
\usepackage{longtable}
\usepackage{subfigure}
\usepackage{textcase}
\usepackage{fancyhdr}
\usepackage{algorithmic,algorithm}
\graphicspath{{figure/}}
\ifCLASSINFOpdf
\else
\fi

\definecolor{ired}{RGB}{255,0,0}

\usepackage[justification=centering]{caption}
\usepackage{setspace}
\begin{document}
\title{Recent Advances in Rate Control: From Optimisation to Implementation and Beyond}
	\author{Xuekai Wei, Mingliang Zhou$^*$, Heqiang Wang, Haoyan Yang, Lei Chen, and Sam Kwong, \IEEEmembership{Fellow, IEEE}
\vspace{-1em}
\thanks{This work was supported in part by the National Natural Science Foundation of China under Grant 62176027; in part by the General Program of the National Natural Science Foundation of Chongqing under Grant cstc2020jcyj-msxmX0790; in part by the Human Resources and Social Security Bureau Project of Chongqing under Grant cx2020073; in part by the Hong Kong GRF-RGC General Research Fund under Grant 11209819 and Grant CityU 9042816; in part by the Hong Kong GRF-RGC General Research Fund under Grant 11203820 and Grant CityU 9042598; and in part by the Hong Kong Innovation and Technology
Commission, InnoHK Project Centre for Intelligent Multidimensional Data Analysis (CIMDA). ($^*$Corresponding author: Mingliang Zhou.)}
\thanks{Xuekai Wei, Mingliang Zhou, Heqiang Wang, Haoyan Yang, and Lei Chen are with the School of Computer Science, Chongqing University, Chongqing 400044, China (e-mail: xuekaiwei2-c@my.cityu.edu.hk, mingliangzhou@cqu.edu.cn, 202114021070t@cqu.edu.cn, 202114131176@cqu.edu.cn, 202114021026t@cqu.edu.cn).}
\thanks{Sam Kwong is with the Department of Computer Science, City University of Hong Kong, Kowloon 999077, Hong Kong (e-mail: cssamk@cityu.edu.hk).}
}
	\maketitle

\thispagestyle{fancy}
\fancyhead{}
\lhead{\scriptsize IEEE TRANSACTIONS ON CIRCUITS AND SYSTEMS FOR VIDEO TECHNOLOGY, VOL. XX, NO. X, X 2023: DOI: 10.1109/TCSVT.2023.3287561}           
\rhead{\thepage}
\fancyfoot{}
\cfoot{\scriptsize Copyright $\copyright$ 20xx IEEE. Personal use of this material is permitted. However, permission to use this material for any other purposes must be obtained from the IEEE by sending an email to pubs-permissions@ieee.org.\\Digital Object Identifier or DOI: 10.1109/TCSVT.2023.3287561.}
\renewcommand{\headrulewidth}{0pt}
\renewcommand{\footrulewidth}{0pt}

\begin{abstract}
Video coding is a video compression technique that compresses the original video sequence to produce a smaller archive file or reduce the transmission bandwidth under constraints on the visual quality loss. Rate control (RC) plays a critical role in video coding. It can achieve stable stream output in practical applications, especially real-time video applications such as video conferencing or game live streaming. Most RC algorithms either directly or indirectly characterise the relationship between the bit rate (R) and quantisation (Q) and then allocate bits to every coding unit so as to guarantee the global bit rate and video quality level. This paper comprehensively reviews the classic RC technologies used in international video standards of past generations, analyses the mathematical models and implementation mechanisms of various schemes, and compares the performance of recent state-of-the-art RC algorithms. Finally, we discuss future directions and new application areas for RC methods. We hope that this review can help support the development, implementation, and application of RC for new video coding standards.
\end{abstract}

\section{Introduction}
\IEEEPARstart{V}{ideo} coding is a technology for converting uncompressed digital video signals into standardised and decodable formats that consume less space. As the speed supported by internet infrastructures, including fixed broadband, mobile networking and Wi-Fi, has increased, video traffic has come to dominate global data traffic; video transmission currently represents approximately 82\% of all traffic, and this percentage is still rising \cite{w1}.
However, uncompressed videos occupy a large quantity of bits, and this high data volume greatly limits various video applications if suitable coding or compression is not applied \cite{smith2013live}. To address this issue, multiple generations of video coding standards have been successively proposed through collaboration among several international standards-setting organisations, e.g., the International Telecommunication Union Telecommunication Standardization Sector (ITU-T) Video Coding Experts Group (VCEG) and the International Organization for Standardization (ISO)/International Electrotechnical Commission (IEC) Moving Picture Experts Group (MPEG). The most well-known recent standards are the Advanced Video Coding (AVC) standard \cite{avc}, the High Efficiency Video Coding (HEVC) standard \cite{hevc} and the Versatile Video Coding (VVC) standard \cite{vvc}. Throughout the history of video standards, each generation of standards has offered significantly improved rate--distortion performance \cite{berger2003rate}, with the main objective of either reducing the coding bit rate as much as possible while ensuring a certain video quality or reducing the coding distortion as much as possible while maintaining a certain coding bit rate limit.
Usually, the original image is divided into multiple square blocks of pixels \cite{tang2019fast}, which are processed in sequence, followed by intra-frame/inter-frame prediction \cite{huang2011predictive}, transformation and inverse transformation \cite{malvar2003low}, quantisation and inverse quantisation \cite{budagavi2014hevc}, loop filtering \cite{norkin2012hevc}, and entropy coding \cite{sze2012high} to finally obtain a video stream.
\begin{figure*}[!t]
	\centering
	\includegraphics[width=0.95\textwidth]{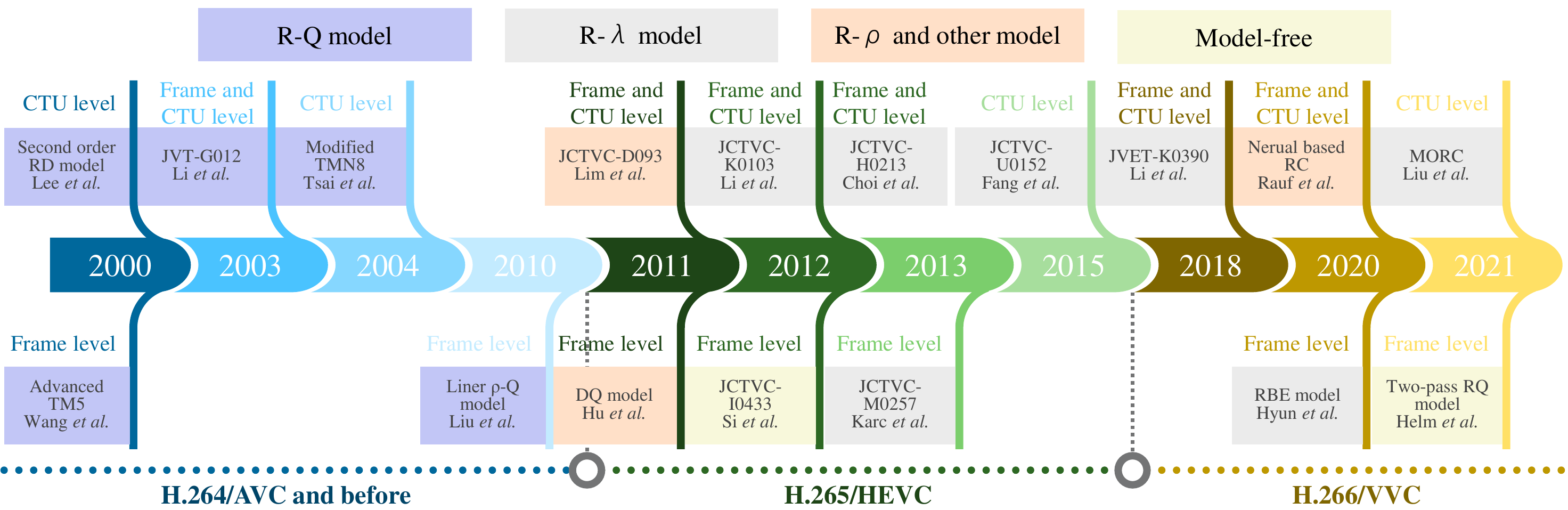}\vspace{-0.5em}
\caption{\centering{History of the main proposals for RC methods.}}
	\label{img1}
\vspace{-2em}
\end{figure*}
\begin{figure}[!t]
\centering
\includegraphics[width=0.3\textwidth]{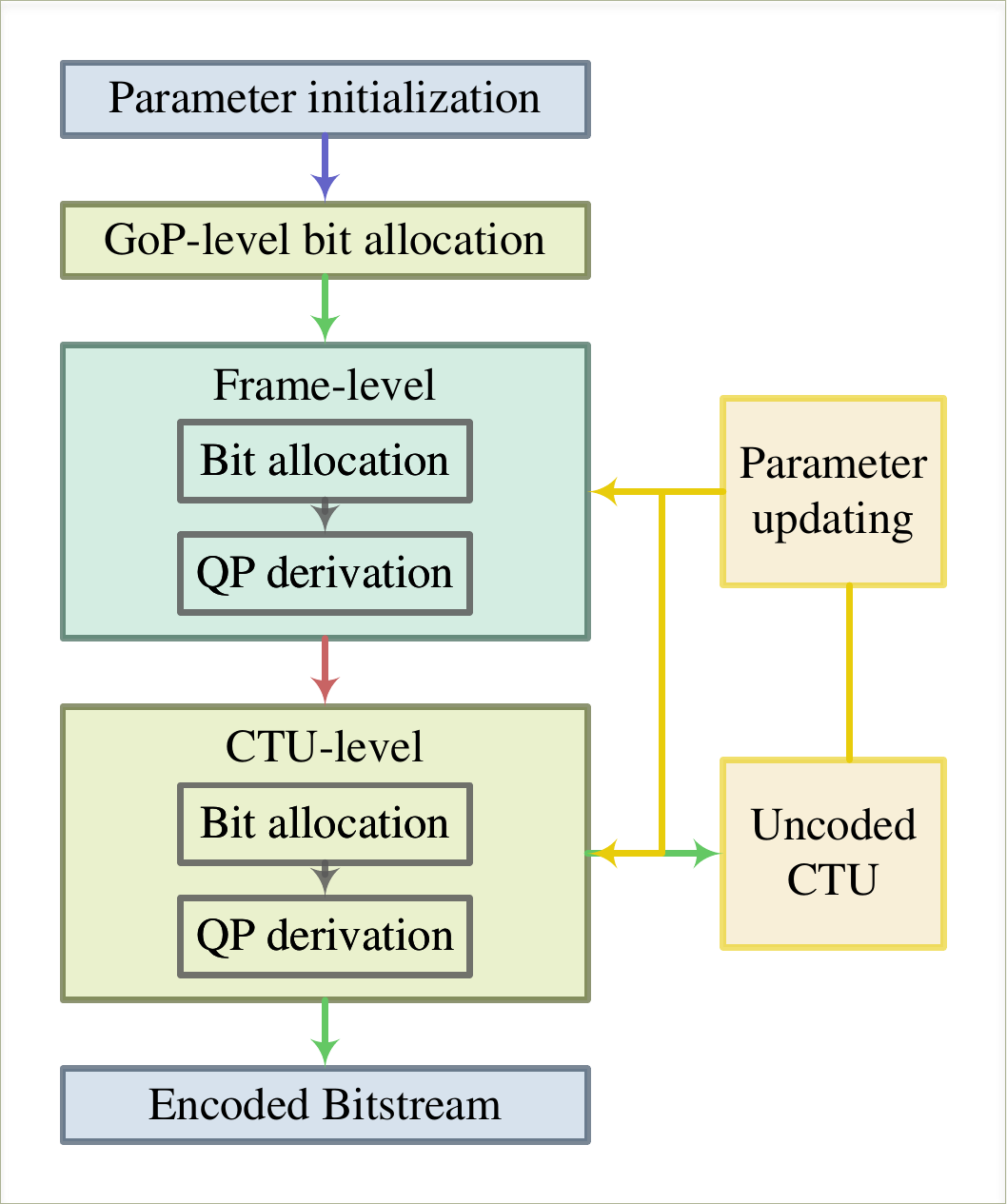}\vspace{-0.5em}
\caption{\centering{Flowchart of the classic RC scheme in HEVC.}}
	\label{img2a}
\vspace{-1.5em}
\end{figure}

Rate control (RC) is a mechanism for determining how many bits are to be transmitted during the encoding process, which is useful because the available bandwidth for video transmission is usually limited. To support the effective transmission of video data while guaranteeing the playback quality of the video service under the condition that all relevant channel bandwidth and transmission delay constraints are met, an RC mechanism selects a series of encoding parameters so as to simultaneously ensure that the bit rate of the to-be-encoded video satisfies the required rate limit and that the encoding distortion is as low as possible. The coding parameters usually include partition models, prediction models and quantisation parameters (QPs). Since a large number of intra-frame and inter-frame prediction techniques are applied in video compression algorithms, the levels of rate--distortion performance achieved for the numerous coding units are interdependent, and the QP of each coding unit is directly determined in accordance with rate--distortion optimisation (RDO) technology. The complexity of obtaining straightforward closed-form solutions for the coding parameters is extremely high. Therefore, an actual RC scheme is usually divided into two steps. First, bits are assigned to each basic coding unit so as to achieve the minimum distortion in accordance with the total bit budget; this process is called bit allocation. Second, in accordance with the relationship model between the coding rate and the QP, the QP is independently determined for each coding unit in accordance with its target number of bits. Because an RC module is a necessary component of any video encoder, all video coding standards have their own recommended RC models, such as TM5 \cite{tm5} of MPEG-2, TMN8 \cite{tmn8} of H.263, JVT-G012 \cite{jvt-g012} of H.264/AVC, JCTVC-H0213 \cite{y71} and JCTVC-K0103 \cite{k0103} of H.265/HEVC, and JVET-K0390 \cite{y47} of H.266/VVC.

From the perspectives of mathematical models and realisation mechanisms, various possible RC schemes have been extensively explored. These schemes have essentially been developed to explore the relationship between rate (R) and distortion (D); that is, different models are based on different R-D curves. Based on the assumption that the encoder can determine the target bit rate by choosing a suitable Q, R-Q-based RC involves estimating the relationship between R and D in the QP domain; hence, these schemes are called Q-domain RC schemes. Many works have followed this idea. Liu \textit{et al.} \cite{c11} proposed an accurate linear R-Q model to characterise the relationship between the total R for both texture and nontexture information and the QP. Hu \textit{et al.} \cite{c14} proposed a frame-level RC algorithm that employs bit information in the RDO process instead of the mean absolute deviation (MAD) of the residuals to predict the complexity of each frame and uses a self-adaptive exponential R-Q model to apply RC. Choi \textit{et al.} \cite{y71} proposed a unified R-Q (URQ) model that can be employed for RC at any level (groups of pictures (GOPs), frames, or basic units) because it captures the relationship between the target rate R and the QP value for a pixel. R-Q-based RC algorithms are widely used in AVC. However, an investigation of JCTVC-I0426 \cite{i0426} has revealed that the slope $\lambda$ of the R-D curve is more important than the QP for bit rate determination. Therefore, Li \textit{et al.} \cite{k0103} proposed a $\lambda$-domain RC method and implemented it in the HEVC standard. Karczewicz \textit{et al.} \cite{y12} proposed an R-$\lambda$-based RC scheme for intra frames/slices based on the sum of absolute transformed differences (SATD) rather than the mean square error (MSE). Fang \textit{et al.} \cite{y14} proposed an R-$\lambda$-based RC method with a preencoding process. There is also another kind of RC algorithm that builds an association between R and the percentage of zeros among the quantised transform coefficients ($\rho$). He \textit{et al.} \cite{w3} estimated an R-D function with low computational complexity in the $\rho$ domain and proposed an encoder-based rate-shape-smoothing algorithm. Liu \textit{et al.} \cite{w4} proposed a new linear model to obtain the QP at the frame level, for which the model parameters in the $\rho$ domain can be adaptively estimated from temporal or interlayer information. Additional milestone papers and studies are also represented in Fig. \ref{img1}. We can observe that R-Q models are applied most frequently in H.264, whereas R-$\lambda$ models are widely adopted in H.265 and H.266, reflecting the superiority of the latter. Another observation is that RC methods are usually implemented at the frame or block level based on considerations regarding the effectiveness and complexity of the encoding process.

\begin{figure}[!t]
\centering
\includegraphics[width=0.411\textwidth]{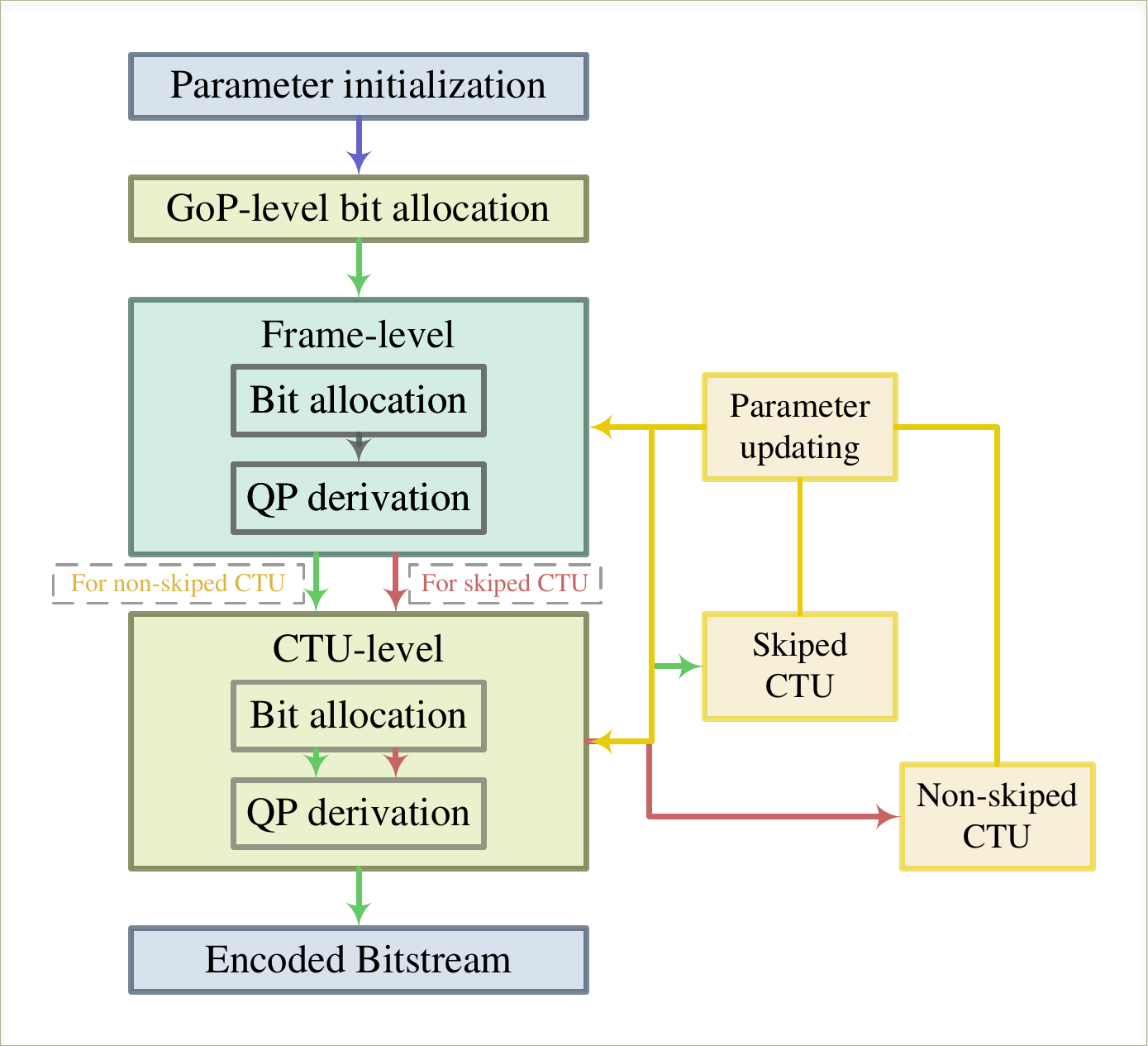}\vspace{-0.5em}
\caption{\centering{Flowchart of the classic RC scheme in VVC. }}
	\label{img2b}
\vspace{-1.5em}
\end{figure}
In this paper, we provide a comprehensive review and survey of RC schemes, from the most widely used methods in engineering to the latest research directions in academia. Moreover, a variety of mathematical models of the relationship between R and Q are introduced and analysed to explore their advantages and disadvantages. Different kinds of RC algorithms are also compared in terms of coding efficiency and time consumption. Finally, recent advances in RC research and future research directions are considered.

The remainder of this paper is organised as follows. Section \uppercase\expandafter{\romannumeral2} introduces and analyses RC models from different perspectives. In Section \uppercase\expandafter{\romannumeral3}, the performance comparison of different RC schemes is discussed. Section \uppercase\expandafter{\romannumeral4} surveys recent and future RC techniques, and Section \uppercase\expandafter{\romannumeral5} concludes the paper.

\section{Brief Overview of Rate Control}
Before we begin exploring recent advances in RC, we introduce the RC structures used in HEVC and VVC. In HEVC, the RC process begins with parameter initialisation; then, bit allocation is performed at three main levels, namely, the GOP level, the frame level and the coding tree unit (CTU) level. The parameters are updated when the features of the coding units change. In particular, QP determination is an important step in RC \cite{hevc}. Compared with HEVC, the skip mode at the CTU level is modified in VVC \cite{vvc}. More details are shown in Fig. \ref{img2a} and \ref{img2b}. This section provides a brief introduction to RC models, divided into six parts: R-Q models, exponential R-$\lambda$ models, R-$\rho$ models, deep learning models, scalable video coding and other types of models.
\begin{table*}[!t]
	\centering
	\caption{\textsc{Representative methods based on R-Q models}}\vspace{-0.4em}
	\label{tab2}
	\setlength{\tabcolsep}{9pt}
	{
		\linespread{0.2}
		\scriptsize
		\begin{tabular}{cccc}
			\toprule[2pt]
			\midrule
			Author & Category & Year & Key idea \\
			\midrule
			Lee \textit{et al.} \cite{c3} & Macroblock & 2000 & \makecell*[c]{Proposed a scalable RC scheme with a more accurate second-order R-D model }\\
			\cline{0-1}	\cline{4-4}
			He \textit{et al.} \cite{c4} & Frame level & 2003 & \makecell*[c]{Proposed an LRC algorithm for the JVT encoder combined with a simple frame-level\\ statistical acquisition scheme }\\
			\cline{0-1}	\cline{4-4}
			Jiang \textit{et al.} \cite{c5} & Frame level & 2004 & \makecell*[c]{Used the MAD ratio as a measure of global frame coding complexity }\\
			\cline{0-1}	\cline{4-4}
			Xu \textit{et al.} & Frame level & 2004 & \makecell*[c]{Assigned different bits to different modes to avoid the regime of poor behaviour of quadratic R-D models }\\
			\cline{0-1}	\cline{4-4}
			Li \textit{et al.} \cite{c7} & Basic unit & 2004 & \makecell*[c]{Taking a macroblock, slice or frame as the basic unit, used a linear model to solve the chicken-and-egg\\ problem for RC in H.264 }\\
			\cline{0-1}	\cline{4-4}
			Ma \textit{et al.} \cite{c8} & \makecell*[c]{Frame level and\\ macroblock} & 2005 & \makecell*[c]{Using a newly proposed R-D model, developed an RC scheme with tuneable complexity}\\
			\cline{0-1}	\cline{4-4}
			Yuan \textit{et al.} \cite{c9} & Frame level &	2006 & \makecell*[c]{Proposed an adaptive coding feature prediction method using spatiotemporal correlation\\ to improve R-D modelling accuracy }\\
			\cline{0-1}	\cline{4-4}
			Kwon \textit{et al.} \cite{c10} & Macroblock &	2007 & \makecell*[c]{Proposed an RC scheme for H.264 video coding using an enhanced R-D model }\\
			\cline{0-1}	\cline{4-4}
			Liu \textit{et al.} \cite{c11} & Macroblock &	2007 & \makecell*[c]{Established an exact linear R-Q model to describe the relationship between the total number\\ of bits of texture and nontexture information and the QP }\\
			\cline{0-1}	\cline{4-4}
			Wang \textit{et al.} \cite{c12} & Macroblock &	2008 & \makecell*[c]{Proposed a joint R-D optimisation RC algorithm for H.264 }\\
			\cline{0-1}	\cline{4-4}
			Tsai \textit{et al.} \cite{c13} & Intra frame &  2010 & \makecell*[c]{Proposed the use of Taylor series rate--QS and scene-change-aware models to determine\\ the QPs of general intra frames and scene-change frames }\\
			\cline{0-1}	\cline{4-4}
			Hu \textit{et al.} \cite{c14} & Frame level &  2010 & \makecell*[c]{Proposed an RC model based on an adaptive exponential R-Q model }\\
			\cline{0-1}	\cline{4-4}
			Hu \textit{et al.} \cite{c15} & Region-based &  2012 & \makecell*[c]{ Studied inter-frame information to objectively divide a frame into regions based on\\ the R-D behaviour of the frame }\\
			\midrule
			Liang \textit{et al.} \cite{y22} & CTU level & 2013 & \makecell*[c]{Proposed an R-Q model for the approximate logarithmic relationship between the rate and QP in HEVC }\\ \cline{0-1}	\cline{4-4}
			Wang \textit{et al.} \cite{y25} & Frame level & 2013 & \makecell*[c]{Proposed a frame-level RC algorithm with the RPS mechanism for HEVC}\\
			\cline{0-1}	\cline{4-4}
			Tian \textit{et al.} \cite{y24} & Frame level & 2014 & \makecell*[c]{Proposed an RCQ algorithm for HEVC intra-frame RC}\\ \cline{0-1}	\cline{4-4}
			Wu \textit{et al.} \cite{y27} & CTU level & 2016 & \makecell*[c]{For LD video coding, proposed an RC scheme for HEVC }\\ \cline{0-1}	\cline{4-4}
			Hosking \textit{et al.} \cite{y23} & Frame level & 2016 & \makecell*[c]{Proposed an improved intra RC method to produce more accurate predictions }\\
			\midrule
			Mao \textit{et al.} \cite{y56} & Frame level & 2020 & \makecell*[c]{Proposed a dependency factor describing the relationship between a reference frame and a frame to be encoded }\\\cline{0-1}	\cline{4-4}
			Helmrich \textit{et al.} \cite{y61} & Frame level & 2021 & \makecell*[c]{Derived two-pass coding parameters based on a new two-step R-Q model}\\
			\midrule
			\toprule[2pt]
		\end{tabular}
	}
		\vspace{-2em}
\end{table*}
\begin{table*}[!t]
	\centering
	\caption{\textsc{Representative methods based on exponential R-$\lambda$ models}}\vspace{-0.4em}
	\label{tab1}
	\setlength{\tabcolsep}{7.5pt}
		\linespread{0.2}
		\scriptsize
		\begin{tabular}{cccc}
			\toprule[2pt]
			\midrule
			Author & Category & Year & Key idea \\
			\midrule
			Wang \textit{et al.} \cite{c1} & Three joint layers & 2009 & \makecell*[c]{Considered optimisation from the GOP level to the CTU level to improve video quality}\\
			\midrule
			Lee \textit{et al.} \cite{y38} & Frame level & 2013 & \makecell*[c]{Incorporated texture and nontexture factors into frame-level RC to build a $\lambda$-domain model}\\\cline{0-1}	\cline{4-4}
			Li \textit{et al.} \cite{y1} & Frame level and CTU level & 2014 & \makecell*[c]{First proposed an exponential R-$\lambda$ model for HEVC }\\
			\cline{0-1}	\cline{4-4}
			Meddeb \textit{et al.} \cite{y19} & Frame level and CU level &
			2014 & \makecell*[c]{Developed an advanced algorithm for video coding that separates the video content into regions\\ of interest and regions of noninterest and increases the bit allocation for regions of interest\\ while maintaining the overall bit rate}\\
			\cline{0-1}	\cline{4-4}
			Yang \textit{et al.} \cite{y44} & Frame level and CTU level & 2014 & \makecell*[c]{Adjusted the R-D model to correct for buffer overflow and underflow issues at low latency}\\
			\cline{0-1}	\cline{4-4}
			Zhou \textit{et al.} \cite{y46} & CTU level & 2014 & \makecell*[c]{Adjusted the R-D model by replacing the distortion calculated based on the MSE with a SSIM-based\\ distortion related to visual quality}\\\cline{0-1}	\cline{4-4}
			Wang \textit{et al.} \cite{y3} & Frame level & 2015 & \makecell*[c]{Took the complexity at the intra-frame level into consideration to design a GRL model}\\
			\cline{0-1}	\cline{4-4}
			Zeng \textit{et al.} \cite{y20} & CTU level & 2015 & \makecell*[c]{Incorporated human visual acuity into the video coding process by separating the input video into\\ perceptually sensitive areas and less perceptually sensitive areas and increasing the bit allocation\\ for perceptually sensitive areas}\\
			\cline{0-1}	\cline{4-4}
			Zhou \textit{et al.} \cite{y2} & CTU level & 2016 & \makecell*[c]{Incorporated video complexity into the model at both the intra-frame and inter-frame levels}\\
			\cline{0-1}	\cline{4-4}
			Li \textit{et al.} \cite{y5} & CTU level & 2016 & \makecell*[c]{Improved the CTU-level $\lambda$-domain model by  optimising bit allocation}\\\cline{0-1}	\cline{4-4}
			Wang \textit{et al.} \cite{y40} & CTU level & 2016 & \makecell*[c]{Designed an RC model for the LD case that can adapt to rapid variations in coding\\ efficiency}\\\cline{0-1}	\cline{4-4}
			Sanchez \textit{et al.} \cite{y15} & Frame level and CU level & 2018 & \makecell*[c]{Proposed a context-based model that incorporates predictive coding technology and uses\\ a piecewise linear function to approximate the R-D curve}\\\cline{0-1}	\cline{4-4}
			Gong \textit{et al.} \cite{y4} & Frame level & 2017 & \makecell*[c]{Considered the influence of temporal layers in the R-$\lambda$ model to ensure that\\ pictures in different time-domain layers are given different levels of importance for prediction}	\\\cline{0-1}	\cline{4-4}
			Perez-Daniel \textit{et al.} \cite{y17} & Block level & 2018 & \makecell*[c]{Developed the $\lambda$-domain model into a multi-R-$\lambda$ model that considers the\\ wider range of luma values found in HDR video}\\\cline{0-1}	\cline{4-4}
			Guo \textit{et al.} \cite{y37} & Frame level & 2018 & \makecell*[c]{Improved the frame-level $\lambda$-domain model by means of optimised bit allocation}\\\cline{0-1}	\cline{4-4}
			Li \textit{et al.} \cite{y21} & CTU level & 2018 & \makecell*[c]{Developed an advanced  $\lambda$-domain model by using an intra-CTU RC scheme that considers\\ the influence of the drift from earlier CTUs to subsequent CTUs in RC}\\\cline{0-1}	\cline{4-4}
			Mir \textit{et al.} \cite{y16} & Frame level & 2018 & \makecell*[c]{Enhanced the $\lambda$-QP relation for HDR video coding by solving the\\ problem of compression performance degradation caused by different coding standards}\\
			\cline{0-1}	\cline{4-4}
			Zhou \textit{et al.} \cite{y39} & CTU level & 2019 & \makecell*[c]{Considered the issue of visual differences in HDR video and built an R-D model based on this\\ issue}\\\cline{0-1}	\cline{4-4}
			Lim \textit{et al.} \cite{y36} & CTU level & 2019 & \makecell*[c]{Proposed a perceptual luminance-adaptive single-loop encoding method }\\\cline{0-1}	\cline{4-4}
			Chen \textit{et al.} \cite{y41} & CTU level & 2019 & \makecell*[c]{Introduced the variance into the R-D model with the aim of minimising the distortion\\ of the CTUs}\\\cline{0-1}	\cline{4-4}
			Zhou \textit{et al.} \cite{y42} & CTU level & 2020 & \makecell*[c]{Adopted a JND factor to build a $\lambda$-domain model that\\ can well describe the distortion of the perceptual field to be used for bit allocation}\\
			\midrule
			Hyun \textit{et al.} \cite{y53} & Frame level & 2020 & \makecell*[c]{Adjusted the R-$\lambda$ model to address textured and nontextured regions simultaneously\\ in VVC}\\\cline{0-1}	\cline{4-4}
			Chen \textit{et al.} \cite{y57} & Frame level & 2020 & \makecell*[c]{Presented an R-$\lambda$ relationship that uses a quadratic R-D model for VVC}\\\cline{0-1}	\cline{4-4}
			Li \textit{et al.} \cite{y60} & Frame level & 2020 & \makecell*[c]{Presented an RC model based on the influence of skip blocks and adjusted the parameter\\ update strategy}\\\cline{0-1}	\cline{4-4}
			Liu \textit{et al.} \cite{y59} & CTU level & 2021 & \makecell*[c]{Researched the relationship between distortion and $\lambda$ to achieve a balance among\\ bit rate, distortion and video quality}\\
			\midrule
			\toprule[2pt]
		\end{tabular}
	\vspace{-2em}
\end{table*}
\subsection{R-Q methods}
The RC algorithms for H.264 adopt a variety of techniques, including the adaptive basic unit layer (ABUL) approach, the fluid traffic model (FTM), a linear MAD model, and a quadratic rate--distortion model \cite{c2}. A hierarchical bit rate control strategy considering the GOP level, the frame level and the basic unit level is adopted. In the Joint Video Team (JVT)'s proposal, the JVT-G012 bit rate control algorithm is adopted. This algorithm introduces the concept of a basic unit, which is used to divide each frame into several basic units. The basic unit may be a macroblock, a row of macroblocks, a field or a frame. In frame-level bit rate control, the target number of bits per frame is allocated based on the network bandwidth, buffer usage, buffer size, and remaining bits. In basic-unit-level bit rate control, the target bits are averaged based on the remaining target bits for the frame. The quadratic R-Q model adopted in the JVT-G012 algorithm for H.264 is as follows:
\begin{equation}
	R=MAD \times\left(X_{1} / Q_{ step }+X_{2} / Q^{2}_{ step }\right)\label{pythagorean1}
\end{equation}
where \textit{R} represents the number of coding bits required by the encoding quantisation coefficient and $ Q_{ step } $ denotes the quantisation step size of the basic units. $ X_{1} $ and $ X_{2} $ are the model coefficients. The MAD is predicted through the following linear prediction model:
\begin{equation}
	MAD_{c b}=a_{1} \times M A D_{p b}+a_{2}\label{pythagorean2}
\end{equation}
where $MAD_{cb}$ and $MAD_{pb}$ denote the MADs for the current basic unit and the corresponding position in the previous frame, respectively, and $a_1$ and $a_2$ are model coefficients, which are updated through linear regression during the processing of the last macroblock of each basic unit.
Following the emergence of quadratic models, the authors of \cite{c3,c7,c9} further developed models of this kind to increase their coding efficiency. Lee \textit{et al.} proposed a scalable RC scheme with a more accurate second-order R-D model. Xu \textit{et al.} first offered a solution to the chicken-and-egg problem between RC and RDO in H.264 and assigned different numbers of bits to different modes accordingly to avoid the regime of poor behaviour of quadratic R-D models. Yuan \textit{et al.} presented an adaptive coding feature prediction method using spatiotemporal correlations to improve the accuracy of R-D modelling. The works in \cite{c4,c8} were both developed based on linear models. He \textit{et al.} presented a linear RC (LRC) algorithm for the JVT encoder combined with a simple scheme for collecting frame-level statistics. Ma \textit{et al.} proposed a new R-D model using the real quantisation step size and, on this basis, proposed an improved RC scheme for the H.264/AVC encoder. With the development of R-Q models, the works in \cite{c5,c10,c14} took the complexity of the frame content into consideration in R-Q modelling. Jiang \textit{et al.} mitigated video distortion due to strong motion or scene changes by using statistics from previously encoded frames to more accurately predict frame complexity. Kwon \textit{et al.} developed an RC algorithm under a constant bit rate constraint for the H.264 baseline profile encoder. Hu \textit{et al.} first adopted a two-stage RC scheme to decouple RDO and RC, then used bit information to predict the frame complexity in the mode decision process for RDO, and finally proposed an adaptive exponential R-Q model for RC. The works in \cite{c12,c15} used adaptive coding methods. Wang \textit{et al.} proposed an R-D-optimised RC algorithm with adaptive initialisation for H.264. In contrast to traditional RC, the area-based RC method proposed by Hu \textit{et al.} can adaptively control the rate in accordance with the content to attain better subjective and objective quality. Tsai \textit{et al.} proposed determining the QPs of general intra frames and scene-change frames by means of a rate--quantisation step size (QS) model and a scene-change-aware model based on Taylor series \cite{c13}.

For HEVC, the work in \cite{y22} proposed a logarithmic model to describe the relation between distortion and rate. Liang \textit{et al.} determined that the rate and QP approximately obey a logarithmic relation and proposed a logarithmic R-Q model for HEVC. 
The works in \cite{y27,y33} mainly considered the low-delay (LD) case. Wu \textit{et al.} proposed an RC scheme for LD video coding considering the temporal prediction structure of HEVC. Si \textit{et al.} proposed frame-level RC schemes for HEVC designed for LD and random access (RA) coding individually. 
Hosking \textit{et al.} presented an enhanced intrasymbol RC method that can produce more accurate predictions and thus reduce the average mismatch rate \cite{y23}. 
Wang \textit{et al.} proposed a frame-level RC algorithm for HEVC based on the reference picture set (RPS) mechanism, leading to specialisation of the QP determination and RPS mechanisms in HEVC, which considerably improved the coding efficiency \cite{y25}. 
The works in \cite{y43,y35,y24} considered linear models to describe the relation between distortion and rate. Yoon \textit{et al.} combined a linear rate model with an R-Q model based on the Cauchy distribution. Si \textit{et al.} modified a linear model to adjust the QP to the q scale. Tian \textit{et al.} proposed a rate--complexity--QP (RCQ) model for HEVC intra-frame RC, which includes linear distortion quantisation as well as exponential R-Q and linear rate--complexity models. 

For VVC, Mao \textit{et al.} presented an RC model based on transform coefficient modelling and derived corresponding R-Q and D-Q models \cite{y56}. Helmrich \textit{et al.} proposed an RC design based on a two-step R-Q model and derived the two-pass encoding parameters \cite{y61}. Representative R-Q models and their corresponding key ideas are summarised in \cref{tab2}, which is sorted by the year of publication and divided into three parts: classic models used in H.264, classic models used in H.265 and classic models used in H.266. The category column of this table indicates at which layer the RC algorithm is applied.
\begin{table*}[htbp]
	\caption{\textsc{Representative methods based on R-$\rho$ models}}\vspace{-0.4em}
	\label{tab4}
	\setlength{\tabcolsep}{14.5pt}
	{
		\linespread{0.5}
		\scriptsize
		\begin{tabular}{cccc}
			\toprule[2pt]
			\midrule
			Author & Category & Year & Key idea \\
			\midrule
			He \textit{et al.} \cite{y6} & Frame level & 2001 & \makecell*[c]{Proposed a frame-level RC algorithm for discrete cosine transform (DCT) video coding}\\ \cline{0-1}	\cline{4-4}
			He \textit{et al.} \cite{y7} & Frame level & 2001 & \makecell*[c]{Proposed a source modelling framework by introducing the new concepts of the characteristic rate curve\\ and rate curve decomposition}\\ \cline{0-1}	\cline{4-4}
			Shin \textit{et al.} \cite{c6} & Frame level & 2004 & \makecell*[c]{Proposed a frame-level RC method for the macroblock mode and bit rate using RDO\\ estimation of the initial QP }\\
			\cline{0-1}	\cline{4-4}			
			Wang \textit{et al.} \cite{y8} & GOP level & 2013 & \makecell*[c]{Introduced a reference atlas to establish a quadtree coding structure and a new reference\\ frame selection mechanism}\\\cline{0-1}	\cline{4-4}
			Wang \textit{et al.} \cite{y28} & Frame level & 2013 & \makecell*[c]{Targeting the LD configuration problem of HEVC, proposed a frame-level RC algorithm in\\ the quadratic $\rho$ domain}\\ \cline{0-1}	\cline{4-4}
			Biatek \textit{et al.} \cite{y26} & CTU level & 2014 & \makecell*[c]{Proposed a more accurate method of capturing the transform coefficients in HEVC based on a\\ mixed Laplacian distribution}\\
			\midrule
			\toprule[2pt]
		\end{tabular}\vspace{-2em}
	}
\end{table*}
\subsection{$\lambda$-domain methods}
The existing research shows that the slope $\lambda$ of an R-D model can be obtained from the hyperbolic R-D function \cite{y1} as follows:
\begin{equation}
	\lambda=-\frac{\partial D}{\partial R}=C K \cdot R^{-K-1} \triangleq \alpha R^{\beta} \label{pythagorean3}
\end{equation}
where $\alpha$ and $\beta$ are parameters related to the video source. Li \textit{et al.} first proposed this type of model for use in HEVC \cite{y11}.
With the development of R-$\lambda$ models, the works in \cite{y12,y3,y2} took the complexity of the frame content into consideration in R-$\lambda$ modelling. In JCTVC-M0257, Karczewicz \textit{et al.} used video content complexity in a model at the intra-frame/slice level. Wang \textit{et al.} considered the complexity at the intra-frame level and designed a gradient-based R-$\lambda$ (GRL) model. Building on the previous developments, Zhou \textit{et al.} added video complexity to a model for both intra frames and inter frames. 
Wen \textit{et al.} \cite{y13} proposed an R-$\lambda$-based RC algorithm with preencoding that improves encoding performance by means of two solutions: one uses only 16x16 coding units (CUs) for preencoding, while the other uses both a hyperbolic function and an exponential function as an improvement to the R-$\lambda$ model. The work in \cite{y4} took the influence of temporal layers into consideration in the R-$\lambda$ model to ensure that pictures in different time-domain layers are given different levels of importance for prediction. 
The authors of \cite{y5,y37} constructed R-D models on the basis of optimised bit allocation. Li \textit{et al.} improved the CTU-level $\lambda$-domain model based on optimised bit allocation, and Guo \textit{et al.} similarly improved the frame-level $\lambda$-domain model by means of optimised bit allocation. Based on Lagrangian multiplier (LM) theory, Wang \textit{et al.} \cite{c1} considered optimisation from the GOP level to the CTU level, thereby improving the video quality. Tang \textit{et al.} \cite{rminorr204} proposed a generalised rate--distortion--$ \lambda $ (R-D-$ \lambda $) optimisation solution for HEVC RC.

The works in \cite{y15,y19,rminorr201} addressed optimised RC in accordance with the video content. Sanchez \textit{et al.} proposed a context-based model that incorporates predictive coding technology and uses a piecewise linear function to approximate the R-D curve. Meddeb \textit{et al.} developed an advanced algorithm for video coding that separates the video content into regions of interest and regions of noninterest and increases the bit allocation for regions of interest while maintaining the overall bit rate. Li \textit{et al.} optimised bit allocation for multiview texture videos based on interview dependency and spatiotemporal correlation. The works in \cite{y16,y17,y39} mainly considered high-dynamic-range (HDR) video. Mir \textit{et al.} enhanced the $\lambda$-QP relation for HDR video coding to solve the problem of compression performance degradation caused by different coding standards. Perez-Daniel \textit{et al.} developed a $\lambda$-domain model into a multi-R-$\lambda$ model that considers the wider range of luma values found in HDR video. Zhou \textit{et al.} considered the issue of visual differences in HDR video and built an R-D model based on this issue. The works in \cite{y20,y36,y42} considered perceptual RC methods. Zeng \textit{et al.} incorporated human visual acuity into the video coding process by separating the input video into perceptually sensitive areas and less perceptually sensitive areas and increasing the bit allocation for perceptually sensitive areas. Lim \textit{et al.} proposed a perceptual luminance-adaptive single-loop encoding method. Zhou \textit{et al.} adopted a just-noticeable distortion (JND) factor to build a $\lambda$-domain model that can well describe the distortion of the perceptual field to be used for bit allocation. Li \textit{et al.} developed an advanced $\lambda$-domain model by using an intra-CTU RC scheme \cite{y21} that considers the influence of the drift from earlier CTUs to subsequent CTUs in RC. Lee \textit{et al.} incorporated texture and nontexture factors into frame-level RC to build a $\lambda$-domain model \cite{y38}. The works in \cite{y40,y41,rminorr203,y44} considered the LD configuration. Wang \textit{et al.} designed an RC model for the LD case that can adapt to rapid variations in coding efficiency. Chen \textit{et al.} introduced the variance into the R-D model with the aim of minimising the distortion of the CTUs. Guo \textit{et al.} presented a $\lambda$-domain frame-level RC scheme. Yang \textit{et al.} adjusted the R-D model to correct for buffer overflow and underflow issues at low latency. Existing RC methods typically optimise the MSE between the distorted image $ Z_{i} $ and the original image $ Z_j $, i.e., the following cost function:
\begin{equation}
\frac{1}{P} \sum_{p=1}^{P}\left(Z_{j}(p)-a Z_{i}(p)-b\right)^{2}
\end{equation}
In this function, the optimal values of $ a $ and $ b $ are computed as:
\begin{equation}
\left\{\begin{array}{l}
a^{*}=\frac{\operatorname{cov}_{Z_{i}, Z_{j}}}{\sigma_{Z_{i}}^{2}} \\
b^{*}=\mu_{Z_{j}}-a^{*} \mu_{Z_{i}}
\end{array}\right.
\end{equation}
where $ \operatorname{cov}_{Z_{i}, Z_{j}} $ is $\frac{1}{P} \sum_{p}\left(Z_{i}(p)-\mu_{Z_{i}}\right)\left(Z_{j}(p)-\mu_{Z_{j}}\right)$. Zhou \textit{et al.} adjusted the R-D model by replacing the MSE-based distortion evaluation in the R-D model with a distortion based on the structural similarity index measure (SSIM), which is related to visual quality \cite{y46}. 

For VVC, the works in \cite{y47,y53,y59} introduced some adjustments to $\lambda$-domain algorithms. Li \textit{et al.} proposed a three-part scheme: splitting skip and nonskip areas at the picture level, changing the update strategy, and modifying the GOP size to 16. Liu \textit{et al.} proposed the use of an adaptive $\lambda$ ratio estimation algorithm. Hyun \textit{et al.} adjusted the R-$\lambda$ model in VVC to address textured and nontextured regions simultaneously. Liu \textit{et al.} researched the relationship between distortion and $\lambda$ to achieve a balance among bit rate, distortion and video quality. 
The works in \cite{y50,y51,y52} made use of a quality dependency factor (QDF) to improve the coding efficiency. Liu \textit{et al.} used QDF-based bit allocation to improve the coding efficiency. Liu \textit{et al.} also proposed an extension of RC to achieve the configuration in JVET-M0600. Ren \textit{et al.} proposed an extension of the QDF to a low frame rate. 
The works in \cite{y57,y60} introduced model modifications in accordance with the coding structure in VVC. Chen \textit{et al.} presented an R-$\lambda$ relationship using a quadratic R-D model for VVC, and Li \textit{et al.} presented an RC model based on the influence of skip blocks to adjust the parameter update strategy. Representative R-$\lambda$ models and their corresponding key ideas are summarised in \cref{tab1}, which is sorted by the year of publication and divided into three parts: classic models used in H.264, classic models used in H.265 and classic models used in H.266. The category column of this table indicates at which layer the RC algorithm is applied. Notably, for RC strategies, the selection of a suitable R-D model is crucial. Packetwise exponential and hyperbolic models are the two most commonly used models to describe the relationship between bit rate and distortion. Studies on HEVC RC have proven the hyperbolic model to be the better option for that standard. However, the R-D relationship has evolved with the advent of the VVC standard and the integration of new coding tools, necessitating a new R-D model to determine the optimal coding QPs.
\begin{table*}[!t]
	\caption{\textsc{Representative methods based on deep learning models}}\vspace{-0.4em}
	\label{tab3}
	\setlength{\tabcolsep}{6.8pt}
	{
		\linespread{0.5}
		\scriptsize
		\begin{tabular}{cccc}
			\toprule[2pt]
			\midrule
			Author & Category & Year & Key idea \\
			\midrule
			Gao \textit{et al.} \cite{y30} & CTU level & 2017 & \makecell*[c]{Extracted features from previous frames and built a more accurate R-D model by means of\\ machine learning; introduced cooperative game theory into bit allocation to increase the\\ coding efficiency and quality (JML RC)}\\ \cline{0-1}	\cline{4-4}
			Hu \textit{et al.} \cite{y29} & CTU level & 2018 & \makecell*[c]{Adjusted the QP by balancing the relationship between texture complexity and coding rate to obtain\\ lower distortion at the CTU level with the help of reinforcement learning (RL RC)}\\ \cline{0-1}	\cline{4-4}
			Zhou \textit{et al.} \cite{y31} & Frame level and CTU level & 2020 & \makecell*[c]{Adjusted the QP via a deep neural network when processing dynamic video sequences to reduce distortion\\ and bit rate fluctuations}\\ \cline{0-1}	\cline{4-4}
			Wei \textit{et al.} \cite{y32} & Tile level & 2021 & \makecell*[c]{Integrated reinforcement learning and game theory into tile-level bit allocation for 360-degree\\ streaming to increase the quality and coding efficiency}\\
			\midrule
			Raufmehr \textit{et al.} \cite{y54} & Frame level & 2020 & \makecell*[c]{Proposed a video bit rate controller that completely conforms to the constraints of real-time applications }\\\cline{0-1}	\cline{4-4}
			Farhad \textit{et al.} \cite{y55} & GOP level & 2021 & \makecell*[c]{Presented a nonlinear relationship by using a neural network to balance the relationship\\ among the bit rate, buffer size and QP}\\\cline{0-1}	\cline{4-4}
			Wang \textit{et al.} \cite{y58} & Initial intra frame & 2021 & \makecell*[c]{Extracted four highly descriptive features to capture the relationship between the video content and\\ the R-D model }\\
			\midrule
			\toprule[2pt]
		\end{tabular}
	}
		\vspace{-1em}
\end{table*}
\begin{table*}[htbp]
	\centering
	\caption{\textsc{Representative methods based on SVC models}}\vspace{-0.4em}
	\label{tab5}
	\setlength{\tabcolsep}{16pt}
	{
		\linespread{0.2}
		\scriptsize
		\begin{tabular}{cccc}
			\toprule[2pt]
			\midrule
			Author & Category & Year & Key idea \\
			\midrule
			Xu \textit{et al.} \cite{c16} & Spatial layer & 2005 & \makecell*[c]{Proposed performing RDO for temporal subband image coding only on low-pass subband\\ images while applying RC to each spatial layer individually }\\
			\cline{0-1}	\cline{4-4}
			Liu \textit{et al.} \cite{c17} & Base layer & 2008 & \makecell*[c]{Predicted the MAD of the residual texture using the available MAD information from the previous\\ frame in the same layer and the base layer for the same frame }\\
			\cline{0-1}	\cline{4-4}
			Pitrey \textit{et al.} \cite{c18} & Inter layer & 2009 & \makecell*[c]{Proposed a simple and attractive bit rate modelling framework in the $\rho$ domain }\\
			\midrule
			Hu \textit{et al.} \cite{y34} & Frame level & 2011 & \makecell*[c]{Proposed a frame-level RC algorithm based on a linear R-Q model and a linear D-Q model }\\
			\midrule
			\toprule[2pt]
		\end{tabular}
	}\vspace{-1.5em}
\end{table*}

\subsection{Models in the $\rho$ domain}
The percentage of zero coefficients, $\rho$, increases monotonically with the QP when the distribution of the transform coefficients is known, meaning that there is a one-to-one correspondence between $\rho$ and the QP. Hence, the relationship between R and the QP can be established through $\rho$. Shin \textit{et al.} modelled the rate--$\rho$ and QP--$\rho$ relationships and adopted a linear approximation scheme to model the rate--$\rho$ relationship \cite{c6}. Experiments have shown that the $\rho$ of the quantised transform coefficients has a good linear relation with the bit rate R \cite{y6} \cite{y7}:
\begin{equation}
	R(\rho)=\theta(1-\rho)\label{pythagorean4}
\end{equation}
where $ \theta $ is a constant and $\rho$ is a model parameter that is related to the video content. The authors of \cite{y8} argued that a Laplacian distribution is not sufficiently precise to capture the true distribution arising from a quadtree prediction structure. Therefore, a mixed Laplacian distribution was applied to describe this distribution. The authors of \cite{y26} proposed an RC algorithm of decreased complexity in the $\rho$ domain, which predicts the encoding parameters for each collocated CTU. Wang \textit{et al.} proposed a more accurate mixed Laplacian distribution to capture the transform coefficients in HEVC \cite{y28}. Representative $\rho$-domain models and their corresponding key ideas are summarised in \cref{tab4}, which is sorted by the year of publication. The category column of this table indicates at which layer the RC algorithm is applied.

\subsection{Classic methods in deep learning }
Deep learning is an effective approach for solving decision-making problems, and thus, it has recently attracted great interest in the video coding community. Hu \textit{et al.} \cite{y29} adjusted the QP by balancing the relationship between texture complexity and coding rate to achieve lower distortion at the CTU level. Gao \textit{et al.} \cite{y30} extracted features from previous frames and built a more accurate R-D model using machine learning. Cooperative game theory has also been introduced into the bit allocation process to increase coding efficiency and quality. Zhou \textit{et al.} \cite{y31} adjusted the QP via a deep neural network when processing dynamic video sequences to reduce distortion and bit rate fluctuations. Wei \textit{et al.} \cite{y32} integrated reinforcement learning and game theory into tile-level bit allocation for 360-degree streaming to increase the quality and coding efficiency.

For VVC, Li \textit{et al.} presented a convolutional neural network (CNN)-based R-$\lambda$ RC approach for intra-frame coding \cite{y49} that reuses the $\lambda$-domain model used for inter-frame RC in the VVC Test Model (VTM) and trained a CNN to simultaneously predict the two model parameters, alpha and beta. Using a multilayer perceptron (MLP) neural network \cite{y54}, Raufmehr \textit{et al.} presented a video bit rate controller that completely conforms to the constraints of real-time applications. Farhad \textit{et al.} \cite{y55} presented a nonlinear relationship by using a neural network to balance the relationship among the bit rate, buffer size and QP. Wang \textit{et al.} designed an RC algorithm \cite{y58} that extracts four highly descriptive features to capture the relationship between the video content and the R-D model. Representative deep learning models and their corresponding key ideas are summarised in \cref{tab3}, which is sorted by the year of publication and divided into two parts: classic models used in H.265 and classic models used in H.266. The category column of this table indicates at which layer the RC algorithm is applied. With the development of deep learning, Lu \textit{et al.} presented the first end-to-end deep video compression model \cite{y64} that jointly optimises all components for video compression.
\begin{table*}[!t]\small
	\centering
	\caption{\textsc{Representative methods based on other models}}\vspace{-0.4em}
	\label{tab6}
	\setlength{\tabcolsep}{9.5pt}
	{
		\linespread{0.2}
		\scriptsize
		\begin{tabular}{cccc}
			\toprule[2pt]
			\midrule
			Author & Category & Year & Key idea \\
			\midrule
			Jing \textit{et al.} \cite{c19} & Intra frame & 2008 & \makecell*[c]{Proposed a method for selecting accurate QPs for intra-coded frames based on a target bit rate; by considering a\\ gradient-based frame complexity measure, the model parameters can be adaptively updated }\\
			\cline{0-1}	\cline{4-4}
			Liu \textit{et al.} \cite{c20} & 4$ \times $4 block level & 2009 & \makecell*[c]{Focusing on RC at the frame and macroblock levels, 
proposed adjusting the number of bits allocated to each frame\\ and each macroblock in accordance with the motion saliency}\\
			\cline{0-1}	\cline{4-4}
			Shen \textit{et al.} \cite{c21} & Macroblock &  2013 & \makecell*[c]{Proposed allocating more bits to visually important macroblocks at the frame level and conversely allocating\\ fewer bits to unimportant macroblocks }\\
			\midrule
			\toprule[2pt]
		\end{tabular}
	}
	\vspace{-1.5em}
\end{table*}

\subsection{Classic methods in scalable video coding}
The aim of scalable video coding (SVC) is to allow partial streams to be obtained on the decoding side while encoding the video signal only once. Three types of scalability, namely, temporal scalability, spatial scalability and quality scalability, are desired to meet different application requirements in terms of rate or resolution. Hu \textit{et al.} introduced a frame-level RC method based on a linear R-Q model and a linear D-Q model \cite{y34}; the formulation can be expressed as follows:
\begin{equation}
	\frac{Q_{s t e p}^{i}{ }^{2} \theta_{i} \gamma_{i}}{k_{i} m_{i}}=\lambda, i=0, \ldots, N\label{pythagorean5}
\end{equation}
where $m_i$ is the predicted MAD of the remaining frames at level $i$ and $k_i$, $\sigma_i$, and $\gamma_i$ are parameters. Xu \textit{et al.} introduced an effective RC method for a scalable video model (SVM) that inherits features of the sophisticated hybrid RC schemes of JVT \cite{c16}. Liu \textit{et al.} proposed a switched model for predicting the MAD of the residual texture using the MAD information available from the previous frame in the same layer \cite{c17}. Pitrey \textit{et al.} presented a new RC scheme for SVC based on a simple yet attractive bit rate modelling framework in the $\rho$ domain \cite{c18}. Representative SVC models and their corresponding key ideas are summarised in \cref{tab5}, which is sorted by the year of publication and divided into two parts: classic models used in H.264 and classic models used in H.265. The category column of this table indicates at which layer the RC algorithm is applied.

\subsection{Other models}
\subsubsection{Segmented R-Q model}
Based on the understanding that the transformation coefficients obey a Laplace distribution with $\sigma^{2}$, the following segmented model can be obtained \cite{y10}:
\begin{equation}
	R(Q)=\left\{\begin{array}{ll}
		\frac{1}{2} \log _{2}\left(2 \mathrm{e}^{2} \cdot \frac{\sigma^{2}}{Q^{2}}\right), & \frac{\sigma^{2}}{Q^{2}}>\frac{1}{2 \mathrm{e}} \\
		\frac{\mathrm{e}}{\ln 2} \cdot \frac{\sigma^{2}}{Q^{2}}, & \frac{\sigma^{2}}{Q^{2}} \leqslant \frac{1}{2 \mathrm{e}}
	\end{array}\right.
\end{equation}
where $\sigma / Q^{2}>1 / 2 \mathrm{e}$ corresponds to the high-bit-rate situation and $\sigma / Q^{2} \leqslant 1 / 2 \mathrm{e}$ corresponds to the low-bit-rate situation.
\subsubsection{D-Q model}
Seo \textit{et al.} proposed a D-Q model to determine the target distortion for QP generation, as follows \cite{y45}:
\begin{equation}\small
	\begin{aligned}
		D&=\alpha \sum_{i=0}^{N_{\text {Depth }}} w_{i}\Bigg \{ \left(1-P_{i, S}\right)\\
		&\left(\frac{2}{\lambda_{i, \mathrm{NS}}^{2}}+\frac{2 Q}{\lambda_{i, \mathrm{NS}}\left(e^{-\frac{1}{2} \lambda_{i, \mathrm{NS}} Q}-e^{\frac{1}{2} \lambda_{i, \mathrm{NS}} Q}\right)}\right)+P_{i, S} \frac{2}{\lambda_{i, S}} \Bigg \}
	\end{aligned}\label{pythagorean6}
\end{equation}
where \textit{D} is the MSE for a frame, $\alpha$ is a model parameter that compensates for the difference between the actual and estimated distortions, $P_{i,S}$ is the proportion of CUs at depth $i$ for which the skip mode is adopted, \textit{Q} is the \textit{q} step size, $N_{Depth}$ is the maximum CU depth, $\omega_i$ is a weighting factor, and $\lambda_i$ denotes the model parameter for the $i$-th CU depth.

\subsubsection{Adaptive R-Q model}
The method proposed by Jing \textit{et al.} aims to select accurate QPs for intra-coded frames in accordance with a target \textit{R}. The parameters of the model can be adaptively adjusted by considering a gradient-based frame complexity measure \cite{c19}.
\subsubsection{Two-pass methods}
The general idea of two-pass RC is to further optimise the QP of each frame in the second encoding pass in accordance with scene complexity statistics computed in the first pass. For AVC, Lie \textit{et al.} \cite{wlie2005two} proposed performing frame-level rate allocation in the second pass using content-aware models constructed in the first pass. For HEVC, Wang \textit{et al.} \cite{7340805} proposed a SSIM-inspired two-pass RC scheme. The algorithm proposed by Ma \textit{et al.} \cite{c8} consists of a one-pass process and a partial two-pass process at the frame and block levels. Zupancic \textit{et al.} \cite{rminorr202} further proposed a two-pass RC method targeting quality improvement for ultrahigh-definition television (UHDTV) delivery. For VVC, Helmrich \textit{et al.} proposed an RC design based on a two-step R-Q model and derived the two-pass encoding parameters \cite{y61}.
\subsubsection{Visual attention models}
Liu \textit{et al.} utilised the mechanism of human visual attention to guide the RC process by incorporating an attention model of motion. They calculated multilayer saliency maps of motion, which were used to adjust the frame-level bit allocation, resulting in quality improvement \cite{c20}. Shen \textit{et al.} proposed an innovative \textit{R} method that considers human visual attention, in which the stronger the local motion attention is, the greater the \textit{R} that is assigned to the frame \cite{c21}.
\subsubsection{Multithreaded coding methods}
At the frame level, multicore systems with a small number of cores can take advantage of multithreaded coding capabilities, e.g., parallelisation, especially when there is little or no dependency between images, as in the case of images in the same temporal layer or even intra frames \cite{meenderinck2009parallel}. In such cases, parallelisation is simple to implement and incurs relatively small coding efficiency losses. However, the gain that can be achieved through frame-level parallelisation is limited by the GOP size, and such processing increases latency despite improving the processing frame rate \cite{ling2013efficiency}. In addition to frame-level parallelisation, slice-level parallelisation can be another way to improve performance \cite{roitzsch2007slice}. Slices partitioned within a picture are independent of each other \cite{blumenberg2013adaptive,koziri2018combining}, apart from potential in-loop filtering dependencies that may exist at the slice boundaries. Slices need not be associated with each other during the execution of most processes performed during coding, such as prediction and transformation, which are applied across slices, and slice-level parallelisation can enable a dramatic enhancement in coding efficiency due to motion prediction processing \cite{karczewicz2021vvc}. Nevertheless, among the RC methods discussed in this article, almost none use the slice-level RC approach; therefore, we do not present experimental statistics on it. A more fine-grained, block-level parallelisation technique is most widely applied \cite{van2003mapping,alvarez2012improving,clare2012hevc}. However, this approach is more difficult to implement because block-level parallelisation requires a more elaborate scheduling algorithm to ensure the correct ordering of the macroblocks due to their multiple spatial dependencies. Furthermore, wavefront parallel processing (WPP) is a commonly used intra-frame parallelisation method, although it has the drawbacks of limitations on the parallelism that can be achieved and unbalanced computational complexity for RC optimisation. To overcome these drawbacks, Joose \textit{et al.} \cite{WWP01} proposed two real-time RC algorithms with parallelisation techniques, a $\lambda$-domain (LD) algorithm and an R-$\lambda$ model (R-LM) algorithm. An adaptive intra-frame parallelisation method was also proposed in \cite{WWP02} that guarantees higher intra-frame parallelism and more accurate control of parallelisation. A hardware architecture strategy was additionally explored in \cite{WWP03} to improve the parallel acceleration of HEVC hardware.

Representative models discussed in this subsection and their corresponding key ideas are summarised in \cref{tab6}, which is sorted by the year of publication. The category column of this table indicates at which layer the RC algorithm is applied.
\begin{table*}[!t]\small
		\centering
		\setlength\tabcolsep{2pt}
		\caption{\textsc{Performance comparison of frame-level RC methods}}\vspace{-0.4em}
		\begin{tabular}{*{8}{c}}
			\toprule[2pt]
			\midrule
Method 	&	 Anchor 	&	 Configuration 	&		 BD-rate (\%) 	&	 BD-PSNR (dB) 	&	 RC accuracy (\%) 	&	 Latency reduction (\%) &  \\
\midrule													
Lee 2013 \cite{y38} 	&	 HM 16.20  	&	 LDP 	&		-3.11	&	0.11	&	89.19	&	 -0.98   \\\midrule
Wang 2013 \cite{y8} 	&	 HM 16.20	&	 LDB/LDP/RA 	&		 -3.13/-4.00/-6.00 	&	 0.11/0.13/0.23 	&	 99.76/99.77/98.44 	&	 -0.34/0.87/2.58  \\\midrule
Wang 2013 \cite{y25} 	&	 HM 16.20 \cite{y71} 	&	 LDB-HE/LDP-HE 	&	-21.91/-12.03 	&	 0.77/0.41 	&	 99.45/99.32 	&	 1.35/1.62  \\\midrule
Xu 2015 \cite{xu2015new} 	&	 HM 16.20 	&	 LDP 	&	-1.5	&	0.14	&	99.33	&	 1.78  \\\midrule
Song 2017 \cite{song2017new} 	&	 HM 16.20 	&	 RA/LDP 	&	-0.20/-0.20 	&	 0.02/0.02 	&	 99.97/99.98 	&	 2.01/-1.00  \\\midrule
\multirow{2}{*}{Guo 2018 \cite{y37}} 	&	 FWA \cite{y1} 	&	 HM 16.20/LDP/LDB 	&	-4.70/-4.30 	&	0.21	&	 99.93/99.94 	&	 3.22/3.13\\\cline{2-8}
&	 AWA \cite{li2016lambda} 	&	 HM 16.20/LDP/LDB 	&	-2.90/-3.20 	&	0.13	&	 99.92/99.39 	&	 2.97/2.66\\\midrule
Hyun 2020 \cite{y53} 	&	 HM 16.20 	&	 AI/LDB/RA 	&	-0.30/-0.20/-0.30 	&	 0.02/0.01/0.02 	&	 99.89/99.50/86.33 	&	 0.22/0.68/0.91  \\\midrule
Raufmehr 2021 \cite{y54} 	&	 VTM 17.0 \cite{y47} &	 LDB 	&	-3.05	&	0.11	&	99.63	&	 36.61  \\\midrule

			\toprule[2pt]
		\end{tabular}\vspace{-1em}
		\label{tab7}
	\end{table*}
\begin{table*}\small
 		\centering
 		\setlength\tabcolsep{2pt}
 		\caption{\textsc{Performance comparison of block-level RC methods}}\vspace{-0.4em}
 		\begin{tabular}{*{7}{c}}
 			\toprule[2pt]
 			\midrule
Method 	&	 Anchor  	&	 Configuration 	&		 BD-rate (\%) 	&	 BD-PSNR (dB) 	&	 RC accuracy (\%) 	&	 Latency reduction (\%)   \\
\midrule													
Kwon 2007 \cite{c10} 	&	JM 9.4	&	 Li \cite{c2}/LDP 	&	-5.40 		&		0.35 		&	97.29 		&		2.50  \\\midrule
Liu 2007 \cite{c11} 	&	 JM 9.4 	&	 JVT-G012 \cite{jvt-g012}/LDP 	&	-5.10 		&		0.33 		&	90.22 		&		3.07  \\ \midrule
Wang 2008 \cite{c12} 	&	JM 9.4	&	 JVT-G012 \cite{jvt-g012}/LDB 	&	-0.63 		&		0.41 		&	93.47 		&		2.10  \\\midrule
Liu 2009 \cite{c20} 	&	JM 9.4	&	 JVT-G012 \cite{jvt-g012}/LDP 	&	-0.12 		&		0.07 		&	95.33 		&		2.80  \\\midrule
Li 2016 \cite{y5} 	&	HM 16.20	&	 Li \cite{y1} 	&	-5.10 		&		0.15 		&	93.57 		&		0.50  \\\midrule
Chen 2019 \cite{y41} 	&	HM 16.20 &	 LDP 	&	-3.30 		&		0.10 		&	97.27 		&		0.10 \\\midrule
\multirow{3}{*}{Zhou 2019 \cite{y39}} 	&	\multirow{3}{*}{HM 16.20}	&	 LDB 	&	-4.60 		&		5.40 		&	97.56 		&		2.90  \\\cline{3-7}
	&		&	 LDP 	&	-1.20 		&		1.00 		&	97.38 		&		1.84  \\\cline{3-7}
	&		&	 RA 	&	-2.00 		&		1.90 		&	99.87 		&		2.50  \\\midrule
\multirow{3}{*}{Zhou 2020 \cite{y42}} 	&	\multirow{3}{*}{HM 16.20}	&	LDB 	&	-1.35 		&		0.01 		&	97.08 		&		0.01  \\\cline{3-7}
	&	  	&	 LDP 	&	-3.30 		&		0.11 		&	94.14 		&		0.07  \\\cline{3-7}
	&	  	&	 RA 	&	-2.75 		&		0.09 		&	94.89 		&		0.77  \\\midrule
 			\toprule[2pt]
 		\end{tabular}\vspace{-1em}
 		\label{tab8}
 	\end{table*}
\begin{table*}[!t]\small
 		\centering
 		\setlength\tabcolsep{2pt}
 		\caption{\textsc{Performance comparison of joint-level RC schemes}}\label{tab9}\vspace{-0.4em}
 		\begin{tabular}{*{8}{c}}
 			\toprule[2pt]
 			\midrule
Method 	&	 Anchor 	&	 Configuration 	&		 BD-rate (\%) 	&	 BD-PSNR (dB) 	&	 RC accuracy (\%) 	&	 Latency reduction (\%)  & \\
\midrule													
Xu 2005 \cite{c16} 	&	 SVM3.0 \cite{y66} 	&	Default	&	-7.20 	&	1.50 	&	99.50 	&	 -3.02 \\ \midrule
\multirow{2}{*}{Ma 2005 \cite{c8}} 	&	 TM5 	&	Default	&	-2.30 	&	0.20 	&	96.50 	&	 -2.22\\\cline{3-8}
	&	 AVC-TM \cite{y69} 	&	Default 	&	-2.20 	&	0.20 	&	96.70 	&	 -2.21\\\midrule
Wang 2009 \cite{c1} 	&	 JVT-G012 \cite{jvt-g012}  	&	 LDP 	&	-4.20 	&	0.98 	&	95.49 	&	 -0.45 \\ \cline{3-8}
Li 2014 \cite{y1} 	&	  HM 16.20   	&	 LD(noH)/LD(H)/RA 	&	- 3.10/-5.50/-8.90 	&	 0.29/0.55/1.08 	&	 99.94/99.90/99.80 	&	 -3.10/0.10/-3.20  \\\midrule
Yang 2014 \cite{y44} 	&	  HM 16.20  	&	 LDB 	&	-2.20 	&	0.31 	&	99.92 	&	 1.30  \\\midrule
\multirow{2}{*}{Zhou 2020 \cite{y31}} 	&	 Hu \textit{et al.} \cite{y29} 	&	 HM 16.20/AI 	&	-5.00 	&	0.50 	&	99.95 	&	 -114.00 (train)/-1.80 (no train)  \\\cline{3-8}
	&	 Gao \textit{et al.} \cite{y30} 	&	 HM 16.20/LDB 	&	-5.20 	&	0.50 	&	99.86 	&	 -1048.20 (train)/-1.20 (no train) \\\midrule
Sanchez 2018 \cite{y15} 	&	  HM 16.20  	&	 Default 	&	-1.00 	&	0.01 	&	99.93 	&	 -1048.20 (train)/-1.20 (no train)  \\\midrule
 			\toprule[2pt]
 		\end{tabular}
 		\vspace{-1.5em}
 	\end{table*}

\section{Performance Comparison of Different RC Schemes}
The bases for measuring the advantages and disadvantages of an RC algorithm do not solely concern how many bits are saved and how much the visual quality is improved; coding/decoding complexity is also an essential consideration for some real-time video applications. Modern encoders, such as AVC, HEVC and VVC, are designed following a framework of square-block-based hybrid coding, which provides the opportunity for performance gains on machines with multithreading capabilities and even multicore processors \cite{hwang1993advanced,bryant2003computer,ranger2007evaluating}. Usually, the performance gain is calculated as the execution time of the improved algorithm divided by the execution time of the original algorithm. {\color{black}Many related techniques} have been proposed in recent video coding standards, some of which are mentioned in the following. All comparisons were performed under fair and well-controlled test conditions. In the evaluation, the general sequences from each class defined in the Common Test Conditions (CTC) \cite{CTC} were chosen, with possible coding QPs of [22, 27, 32, 37]. The total number of test sequences in each class is as follows: Class A1 contains three sequences, Class A2 contains three sequences, Class B contains five sequences, Class C contains four sequences, Class D contains four sequences, and Class E contains three sequences \cite{CTC}. The resolution of the sequences in each class is as follows: Class A1 has a resolution of 3840$ \times $2160, Class A2 has a resolution of 3840$ \times $2160, Class B has a resolution of 1920$ \times $1080, Class C has a resolution of 832$ \times $480, Class D has a resolution of 416$ \times $240, and Class E has a resolution of 1280$ \times $720 \cite{CTC}. The total number of frames per sequence is taken to be 300, and for sequences that are fewer than 300 frames, the results are normalised accordingly with respect to the actual number of frames. The resolution of the video sequences varies from the Common Intermediate Format (CIF) to 4K, and the bit depth is 8 bits. The target bit rate was set to the actual bit rate obtained by compressing the same sequence at a fixed QP in the non-RC encoding mode. The detailed GOP types used for the experiments can be found in the evaluation tables. For the AVC, HEVC, and VVC RC methods, all other settings were kept the same as the defaults in JM 9.4, HM 16.20, and VTM 17.0, respectively.

Some terms that appear in the tables in this section may need explanation. The \textit{configuration} column represents part of the test conditions \cite{jvt-h1100} recommended by the standardisation group. Its entries represent three different prediction structures: low delay with B frames (LDB), low delay with P frames (LDP) and random access (RA). Some entries have suffixes of \textit{H} or \textit{noH}, which indicate whether bits are allocated hierarchically. Coding efficiency is evaluated in terms of the Bj\o{}ntegaard-delta bit rate (\textit{BD-rate}) \cite{bjontegaard2001calculation}, which indicates the reduction in bit rate at a given quality, while the Bj\o{}ntegaard-delta peak signal-to-noise ratio (\textit{BD-PSNR}) complementarily represents the video quality enhancement at a given bit rate. The results in the \textit{Latency reduction} column, which measure the coding complexity of each algorithm, are normalised to the average time taken to encode a frame. Finally, the RC accuracy, ${Acc}$, is an important index for an RC algorithm that represents the difference between the actual bit rate and the target bit rate and is calculated as follows:
\begin{equation}
	{Acc}=1 - \dfrac{|R_{r}-R_{c}|}{R_{c}}\times 100\%
\end{equation}
where $R_{r}$ and $R_{c}$ are the actual number of bits and the target number of bits, respectively.
\subsection{Implementations at the frame level}
In the vast majority of cases, the working environment of the encoder is the CPU; accordingly, many optimisations have been analysed and implemented in modern CPUs. Thus, unless otherwise mentioned, the following experiments were carried out on the CPU. More detailed quantitative comparisons are shown in \cref{tab7}. When using a CPU, the key to improving performance lies in reducing data dependencies and operational dependencies to improve parallelisation, and at the frame level, these requirements are naturally satisfied. To build an accurate R-Q model, Lee \textit{et al.} \cite{y38} separately constructed three Laplacian probability models for low-textured, medium-textured, and high-textured CUs; the BD-rate of this method is -3.11\%, and the BD-PSNR is 0.11 dB (LDP). Wang \textit{et al.} \cite{y8} considered new coding tools for use in HEVC and adopted a hierarchical RC architecture to maintain the video quality of keyframes. In comparisons considering all three configurations, the BD-rate ranges from -3.13\% to -6.00\%, with corresponding BD-PSNR values ranging from 0.11 dB to 0.23 dB. These authors also proposed D-Q and R-Q models for finding the interframe dependency between to-be-coded frames and the reference frame. Accordingly, a mixed Laplacian distribution $\rho$-domain-based rate--GOP model was proposed. Wang \textit{et al.} \cite{y25} proposed an efficient hierarchical bit allocation scheme based on a new mechanism for HEVC, i.e., the RPS mechanism, which achieves the highest BD-rate among the compared methods. Moreover, they proposed an innovative header bit ratio prediction method to improve RC accuracy and used a quadratic R-Q model to calculate the QP. Xu \textit{et al.} \cite{xu2015new} focused on video sequences depicting discontinuous scenes and proposed a novel bit allocation algorithm by building the correlation between the intensity of a scene change and the bit allocation. A BD-rate of -1.5\% can be achieved in this way, and the BD-PSNR is 0.14 dB. Song \textit{et al.} \cite{song2017new} addressed the issue that the RC method used in AVC is no longer suitable for HEVC due to the differences in the GOP coding structures; accordingly, they proposed a new GOP-level bit allocation method that can achieve more accurate RC and lower bit fluctuation, resulting in slightly better R-D performance, as shown in table \ref{tab7}. Guo \textit{et al.} \cite{y37} considered the temporal R-D dependency in GOP-level bit allocation; on this basis, an advanced frame-level R-D model that can more completely use the information of the coded frames was introduced to further enhance R-D performance. Then, an equation for bit allocation at the frame level was proposed for the optimal Lagrange multiplier approach, which can be solved by employing a recursive Taylor expansion (RTE) scheme. Two comparative experiments were performed using \cite{y1} and \cite{li2016lambda} as benchmarks, referred to as fixed-weight bit allocation (FWA) and adaptive-weight bit allocation (AWA), respectively. The comparison results show that the BD-rate ranges from -4.30\% to -4.70\% for FWA and -2.90\% to -3.20\% for AWA. As an alternative to traditional methods that regard the entire RC process as deterministic, some methods treat the variables and parameters of RC as random variables to re-examine the RC process. Hyun \textit{et al.} \cite{y53} observed that the inaccuracy of existing linear rate estimation models causes a decline in RC performance. Thus, they adopted a method called recursive Bayesian estimation (RBE) to precisely estimate rates. Performance comparisons with all three configurations show that the BD-rate ranges from -0.20\% to -0.30\%, with slight improvements in the BD-PSNR. Raufmehr \textit{et al.} \cite{y54} proposed a video bit rate controller to meet the demands of real-time applications in the VVC standard by suppressing bit fluctuations and buffer overflow and underflow. The BD-rate is -3.05\%, and the BD-PSNR is 0.11 dB. In this method, the necessary QP modifications are estimated by using an MLP neural network at the frame level, which is beneficial for robust buffer control. Comparisons of the results suggest that each method has its strengths and weaknesses, depending on the specific application or situation. Lee \textit{et al.}'s method is beneficial for building an accurate R-Q model, while Wang \textit{et al.}'s method is suitable for maintaining the video quality of keyframes. Xu \textit{et al.}'s method is effective for video sequences depicting discontinuous scenes, while Song \textit{et al.}'s method achieves more accurate RC and lower bit fluctuation. Guo \textit{et al.}'s method considers temporal R-D dependency and can more completely use the information of the coded frames, while Hyun \textit{et al.}'s method can precisely estimate rates. Raufmehr \textit{et al.}'s method is suitable for real-time applications. Due to the elimination of traditional models and the utilised network structure, the model complexity is lower, resulting in a large latency reduction. However, these frame-level methods also have some drawbacks, such as poor scalability or high memory occupation. Fortunately, block-based implementations can address these problems \cite{c24}.

\subsection{Implementations at the block level}
RC evaluation at the block level was carried out using the JM reference software for H.264 and the HM reference software for H.265. In \cref{tab8}, the RC methods proposed by Kwon \textit{et al.} \cite{c10}, Liu \textit{et al.} \cite{c11}, Wang \textit{et al.} \cite{c12} and Liu \textit{et al.} \cite{c20} are all compared and analysed. In the LDP mode of \cite{c10}, the BD-PSNR is improved by 0.35 dB over the baseline. In the scheme of \cite{c11}, the BD-rate is -5.10\%, and the BD-PSNR is 0.33 dB. It can be seen that the algorithm in \cite{c12} can achieve better PSNR results than JVT-G012 \cite{jvt-g012}, while the bit rate performance of the two algorithms is similar. In the LDP mode of \cite{c20}, the BD-rate is -0.12\%, the BD-PSNR is 0.07 dB, and the latency reduction is 2.80\% on average. \cite{y5} proposed a new scheme named the optimal bit allocation (OBA) scheme and presented a detailed comparison based on \cite{y1}; the BD-rate of this method is -5.10\%, and the BD-PSNR is 0.15 dB. Compared with the method of HM 16.20, the average BD-rate of \cite{y41} can reach -3.30\%, and the BD-PSNR can reach 0.10 dB. For the method of \cite{y39}, the BD-rate and BD-PSNR for LDB/LDP/RA are -4.60\%/-1.20\%/-2.00\% and 5.40 dB/1.00 dB/1.90 dB, respectively, which are much higher than those of the baseline HM 16.20 coding algorithms. A recently proposed JND-based \cite{c22, c23} perceptual RC method \cite{y42} is also included in the comparison, and it can be seen that this method effectively reduces the bit rate without compromising the encoded video quality, as confirmed by objective metrics such as the PSNR. In addition, in terms of coding control, the actual \textit{R} after encoding is closer to the target \textit{R}. As seen from these comparisons, different methods are suitable for different situations. The OBA scheme is appropriate for achieving a lower bit rate with good quality. The method of \cite{y39} is suitable for improving the BD-rate and BD-PSNR for LDB/LDP/RA, while the JND-based perceptual RC method is appropriate for reducing the bit rate without compromising the encoded video quality. Other methods also have their own advantages in achieving a higher BD-PSNR or lower BD-rate.

\subsection{Implementations addressing joint layer optimisation}
Often, multiple layers can be jointly optimised to improve the coding efficiency. Recall that the RC process can be divided into two main parts: one is bit allocation, while the other is determining how to realise the target bit allocation for a CU through the R-D model. As shown in table \ref{tab9}, the algorithm proposed by Xu \textit{et al.} \cite{c16} is simultaneously implemented at the GOP, frame and basic unit levels, and experiments in SVM3.0 \cite{y66} show that the mismatch between the target \textit{R} and real \textit{R} does not exceed 0.5\% and that the BD-PSNR on average is approximately 1.50 dB. The algorithm proposed by Ma \textit{et al.} \cite{c8} consists of a one-pass process and a partial two-pass process at the frame and block levels, and experiments show that compared with TM5 and AVC-TM \cite{y69}, the BD-PSNRs are 0.20 dB and 0.20 dB, respectively. The algorithm proposed by Wang \textit{et al.} \cite{c1} is a joint three-layer (JTL) model implemented in JM 9.4 \cite{y68}. Compared with JVT-G012 \cite{jvt-g012}, the average BD-PSNR is 0.98 dB, while the average latency reduction is -0.45\%. The algorithm proposed by Li \textit{et al.} \cite{y1} is implemented at the frame and CTU levels, and testing in HM 16.20 shows that relative to \cite{y71} in the LD(noH), LD(H) and RA configurations, the RC accuracy gains are 0.06\%, 0.10\% and 0.20\%, respectively; the average BD-PSNRs are 0.29 dB, 0.55 dB and 1.08 dB, respectively; and the latency reductions are -3.10\%, 0.10\%, and -3.20\%, respectively. The algorithm proposed by Yang \textit{et al.} \cite{y44} is implemented at the frame and CTU levels, and experiments in HM 16.20 in the LDB configuration show that the bit rate at the same quality is 2.20\% lower than that of the RC algorithm in HM 16.20. The algorithm proposed by Zhou \textit{et al.} \cite{y31} is implemented at the frame and CTU levels, and experiments in HM 16.20 show that compared with \cite{y29} and \cite{y30} under the all-intra (AI) and LDB coding structures, the RC accuracies are 99.95\% and 99.86\%, respectively; the BD-PSNR is 0.50 dB in both cases; and the latency reduction is increased by 114\% and 1048.2\%, respectively, when the time needed to train the model parameters is included and by 1.80\% and 1.20\%, respectively, when the time needed to train the model parameters is not included. The algorithm proposed by Sanchez \textit{et al.} \cite{y15} was also implemented, and experiments in HM 16.20 show that the RC accuracy is 99.93\% with a slight BD-PSNR increase, while the encoding time is increased if the training time is included. The above comparisons indicate that the various algorithms all show improvements in coding efficiency, with different strengths and limitations. Xu \textit{et al.}'s algorithm minimises the mismatch between the target and real \textit{R} values, while Ma \textit{et al.}'s algorithm is effective for one-pass and partial two-pass encoding. A significant PSNR gain along with a slight latency reduction can be guaranteed with Wang \textit{et al.}'s algorithm. Significant gains in RC accuracy and PSNR can also be seen in Li \textit{et al.}'s algorithm. Zhou \textit{et al.}'s and Sanchez \textit{et al.}'s algorithms are suitable for non-real-time encoding solutions.

From the above evaluations, it can be seen that RC modelling methods have evolved over time, with each model having its own set of advantages and disadvantages. The simple and widely used R-Q models cannot accurately reflect the quality for diverse video content. More accurate R-$ \lambda $ models have been proposed to better reflect the R-D relationship, but these models require frequent and complex updates. An R-$ \rho $ model, while improving the linear approximation of the R-D relationship, still establishes only an indirect relationship between R and the QP. Model-free RC methods offer greater flexibility but require more computational resources and have not yet been widely adopted. Overall, RC modelling has become more sophisticated over time, and models should be selected flexibly to balance coding efficiency and performance.
\section{Future Works}
As efficient tools for compressing video information, video codecs allow service providers to compress video files so that they will occupy minimal storage space and can be efficiently delivered over a range of networks. The purpose of bit rate control is to achieve stable and high-quality video compression to the greatest possible extent under specified bit rate constraints. By removing redundant information, RC methods aim to maintain the original video quality while reducing the amount of data sent over the network or occupying storage space. With the development of virtual and augmented reality applications, video users have gained the ability to interact with and influence objects in immersive three-dimensional (3D) simulated environments that emulate reality with the help of interactive devices, thereby providing an experience equivalent to that of an objective natural environment. This situation has driven the further development of video streaming media in the directions of ultrahigh definition, high dynamics, high frame rates, and high depth \cite{rw10,rw02,rw03}.

These developments have resulted in ever-increasing demands on video coding techniques for current applications to achieve higher compression efficiency, lower computational complexity, and more intelligent integration into video analysis systems. We believe that the future directions of related research can be summarised from six perspectives: RC methods based on machine learning and deep learning, development from RC methods to quality control methods, perceptual and depth-aware RC methods, RC methods for various advanced video sources, RDO and RC in depth coding, and point cloud RC methods. Future work is expected to focus on topics such as image integration, video capture, encoding, processing, analysis, and understanding, with the objective of guiding a new generation of codecs to effectively and intelligently model the human visual system. A promising possibility is that through the dynamic selection of different RC models, such as R-Q, R-$ \lambda $, or R-$ \rho $ models, or model-free methods in accordance with different video content or coding conditions, the trade-off between coding efficiency and RC accuracy can be optimised \cite{rw21b}. For instance, an R-$\lambda$ model performs best in terms of coding efficiency and RC accuracy in texture regions of HDR content, but it may not be suitable for nontexture regions. On the other hand, an R-Q model, which is relatively simple, may be more appropriate for simpler motions or textures. By switching between different models based on region characteristics or coding conditions, better RC accuracy and general quality can be achieved while maintaining high coding efficiency. Therefore, more sophisticated algorithms and strategies for dynamically selecting and adapting RC models in real-time video coding applications should be developed. The specific directions of anticipated future development are discussed as follows.

\subsection{Learning-based visually enhanced RC methods}
Machine learning (ML) and deep learning (DL) techniques are widely acknowledged as important tools for analysing and processing massive amounts of weakly correlated or high-dimensional data \cite{rw04,rw06,rw07,rw08,rw09}. As new technologies for video applications (e.g., virtual reality, augmented reality, and point clouds) revolutionise the video coding industry, the heterogeneity and complexity of the captured data are presenting increasing challenges for the efficient compression of these data. Based on the above review of the methods applied to date for video RC in the ML and DL domains, this paper argues that future ML and DL techniques can help to achieve smarter video coding.

Based on the specific requirements of video coding tasks, learning-based RC methods aim to achieve intelligent RC with low complexity, high coding efficiency, and high visual quality. First, ML and DL techniques can be used to combine analysis and recognition tasks with encoding tasks, enabling intelligent RC through the effective reuse of video information and reducing the complexity of video encoding. Second, various learning methods, such as active learning, reinforcement learning, and transfer learning, can be introduced to establish self-updating mechanisms in the relevant models to solve the more complex RC decision-making problems arising in new generations of video coding standards. Third, for the current ML and DL methods applied for video RC, there is still a need to solve the problems of their relatively high computational complexity and cost. Thus, another important direction of future development for ML and DL RC methods will be to investigate how to realise low-hardware and low-cost implementation solutions based on DL.

\subsection{Perceptual and depth-aware RC methods}
Perceptual-based video coding that exploits perceptual redundancy is a promising research area worthy of future consideration \cite{rw13,rw14,rw15}. Related research aims to realise an optimal RC mechanism by constructing the relationship between video quality assessment (VQA) results and the coding rate. When a VQA algorithm is to be applied to determine the quality target in a video coding module, it is necessary to adapt the VQA algorithm from image/video-based assessment to block-based assessment and to rate--distortion theory. A convenient method is to model the relationship between the VQA metric of interest and the MSE, given that the MSE is the basis of RDO. In addition, balancing the uniqueness and particularities of perceptual redundancy is also a fundamental difficulty that needs to be overcome. Since the perception of the human visual system for static, dynamic, stereo, and omnidirectional video varies from person to person, it is a daunting challenge to implement a general VQA algorithm suitable for different applications. In addition, the introduction of VQA makes RC complexity control particularly challenging. The computational complexity increases substantially with the adoption of more advanced feature extraction tools and learning-based classifiers for quality prediction. Frequent calls to learning-based VQA algorithms in intensely learning-based RDO schemes can result in extremely complex encoding algorithms. Thus, it will be worth investigating how to integrate VQA, especially better-performing learning-based VQA, into video coding while maintaining a desirable complexity. Future work on perceptual RC will be aimed at developing an objective visual perception model that is broadly advantageous in terms of accuracy, complexity, and adaptability.

\subsection{Interoperable end-to-end RC methods for hyperrealistic and high-dimensional videos}
The rapid evolution of immersive video applications, such as augmented reality (AR), virtual reality (VR) and HDR videos, is presenting new challenges for end-to-end RC methods. Such immersive applications require calculating the positions and angles of camera images in real time and generating corresponding artificial images, with the aim of simulating or supporting interaction with the real world \cite{rw16,rw17,rw18,rw19,rw20,rw21}. A VR system uses computer simulation to generate a 3D space, presenting users with a visual and interactive simulation without restrictions. A much more advanced technology is mixed reality (MR), which mixes the real-world environment with AR and VR technologies. This unique experience requires redefining the perceptual quality indicators used in the video encoding process, implementing end-to-end RDO, and outputting stable and high-quality code streams while taking viewing interoperability into consideration. Interoperability can be defined as the ability to achieve a stable viewing angle bit rate through interaction during the encoding process, involving the sharing of user viewing data and encoding information between the client and the server through data exchange via the head-mounted device (HMD). RC interoperability for immersive video is closely related to interoperability in broader human--computer interaction. The need for interoperability between immersive video delivery systems increases the practical value of RC coding. This interoperability can be achieved in two ways: through competition for more network bandwidth and by enabling data-driven decisions based on users' viewing habits. The users of such immersive experiences will also benefit from improved data quality and a better immersive interactive experience. Therefore, an important direction of development for hyperrealistic and high-dimensional video technology will be to implement an R-D model based on HDR video characteristics in order to improve coding performance and achieve a globally optimal rate allocation scheme in the subsequent RC mechanism. In addition, video content prediction has become very popular in recent years due to its ability to learn from previous viewing behaviour to construct the forthcoming video. Such predicted video content can be widely used in decision-making, autonomous driving, video comprehension, etc. Investigating how best to achieve RC for this type of video will be very important, as all of the abovementioned related tasks demand smooth and high-quality streaming video.

\subsection{Beyond 5G/6G-powered quality control methods}
With the popularisation of 5G networking, high-data-rate and low-latency network connections are increasingly expected to ensure a smooth and high-quality playback experience for video users \cite{rminorr301,rminorr302,rminorr303}. Since visual quality is an important aspect of the user experience in many media applications, high quality must be guaranteed while pursuing high smoothness. These issues may become critical as the demand for high-quality video transmission becomes more widespread. The objective of quality control is to keep the video quality within a certain high range under the premise of high bandwidth, which can be achieved by using variable bit rates. However, most existing techniques may not provide constant visual quality and/or efficient compression. It will be more critical for future quality control methods to consider the quality difference between frames as a measure of frame complexity in order to model the relationship among the target bit rate, distortion, and QP. Another essential direction of research will be to pursue the implementation of a low-latency quality control scheme to achieve quality stability.

\section{Conclusions}
This paper comprehensively reviews the latest progress in RC techniques for the H.265/HEVC and H.266/VVC video coding standards and discusses relevant development prospects. More specifically, this paper first introduces and compares different kinds of RC methods based on various R-D models as well as emerging DL-based schemes. Then, the implementation schemes of RC methods on different hardware platforms are reviewed. Finally, we discuss RC methods based on ML and DL, the evolution from RC to quality control, RC methods considering human visual perception and depth perception, RC methods for 360-degree/HDR video, and RC methods for predicted video content. A comprehensive summary of directions of future work on topics such as RC methods and RDO in depth coding is also presented. The aim is to provide valuable guidance for the improvement, implementation, application, and continuous development of the current and next generations of RC standards.

\renewcommand{\IEEEbibitemsep}{0pt plus 0.4pt}
\makeatletter
\IEEEtriggercmd{\reset@font\normalfont\fontsize{7.3pt}{8.3pt}\selectfont}
\makeatother
\IEEEtriggeratref{1}


	\begin{IEEEbiography}[{\includegraphics[width=1in,height=1.25in,clip,keepaspectratio]{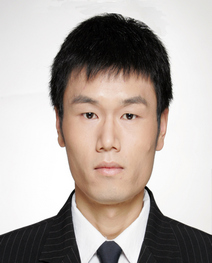}}]{Xuekai Wei}
		received the bachelor's degree in Electronic Information Science and Technology from Shandong University in 2014, the master’s degree in Communication and Information Systems from Shandong University in 2017, and the Ph.D. degree in computer science from the City University of Hong Kong, Hong Kong, China in 2021. He was a post-doctoral at the School of Artificial Intelligence, Beijing Normal University, Beijing, China, from 2021 to 2022. He is currently an Associate Professor with the School of Computer Science, Chongqing University, Chongqing, China. His current research interests include video coding/transmission and machine learning.  
	\end{IEEEbiography} 

	\begin{IEEEbiography}[{\includegraphics[width=1in,height=1.25in,clip,keepaspectratio]{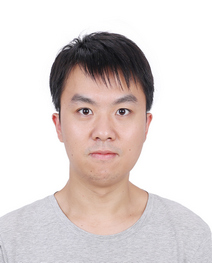}}]{Mingliang Zhou}
		received the Ph.D. degree in computer science from Beihang University, Beijing, China, in 2017. He was a Postdoctoral Researcher with the Department of Computer Science, City University of Hong Kong, Hong Kong, China, from September 2017 to September 2019. He was a Postdoctoral Fellow with the State Key Laboratory of Internet of Things for Smart City, University of Macau, Macau, China, from October 2019 to October 2021. He is currently an Associate Professor with the School of Computer Science, Chongqing University, Chongqing, China. His research interests include image and video coding, perceptual image processing, multimedia signal processing, rate control, multimedia communication, machine learning, and optimization.  
	\end{IEEEbiography} 	
	
	\begin{IEEEbiography}[{\includegraphics[width=1in,height=1.25in,clip,keepaspectratio]{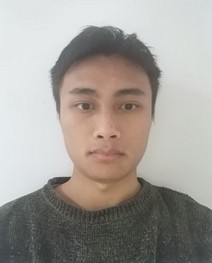}}]{\\Heqiang Wang}
	is currently a M.S. student of Computer Science in Chongqing University. He received his bachelor degree of Mechanical Design Manufacturing and Automation in Southwest University. His research interests include perceptual image quality assessment and video coding.
	\end{IEEEbiography}
	
	\begin{IEEEbiography}[{\includegraphics[width=1in,height=1.25in,clip,keepaspectratio]{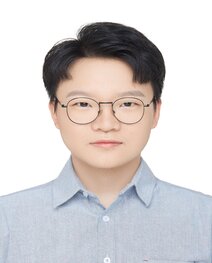}}]{\\Haoyan Yang} 
	is currently a M.S. student of Electronic Information in Chongqing University. He received his bachelor degree of Software Engineering in Nanjing Audit University. His research interests is video coding.
	\end{IEEEbiography}

	\begin{IEEEbiography}[{\includegraphics[width=1in,height=1.25in,clip,keepaspectratio]{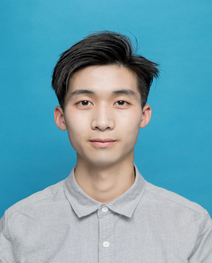}}]{\\Lei Chen}
	is currently a M.S. student of Computer Science in Chongqing University. He received his bachelor degree of Medical Information Engineering in Zhejiang Chinese Medical University. His research interests include image super-resolution and video coding.
	\end{IEEEbiography}

	\begin{IEEEbiography}[{\includegraphics[width=1in,height=1.25in,clip,keepaspectratio]{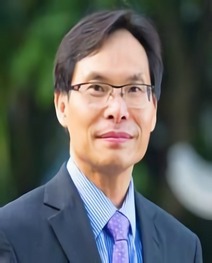}}]{Sam Kwong}
	(M'93-SM'04-F'14) received the B.S. degree from the State University of New York at Buffalo, Buffalo, NY, USA, in 1983, the M.S. degree in electrical engineering from the University of Waterloo, Waterloo, ON, Canada, in 1985, and the Ph.D. degree from the University of Hagen, Hagen, Germany, in 1996.
	
	From 1985 to 1987, he was a Diagnostic Engineer with Control Data Canada, Mississauga, ON, Canada. He joined Bell Northern Research Canada, Ottawa, ON, Canada, as a member of Scientific Staff. In 1990, he became a Lecturer in the Department of Electronic Engineering, The City University of Hong Kong, Hong Kong, China, where he is currently a Professor with the Department of Computer Science. His research interests include video and image coding, and evolutionary algorithms.
	
	Prof. Kwong serves as an Associate Editor for the IEEE Transactions on Industrial Electronics and the IEEE Transactions on Industrial Informatics.
	\end{IEEEbiography} 


\begin{thebibliography}{100}
\providecommand{\url}[1]{#1}
\csname url@samestyle\endcsname
\providecommand{\newblock}{\relax}
\providecommand{\bibinfo}[2]{#2}
\providecommand{\BIBentrySTDinterwordspacing}{\spaceskip=0pt\relax}
\providecommand{\BIBentryALTinterwordstretchfactor}{4}
\providecommand{\BIBentryALTinterwordspacing}{\spaceskip=\fontdimen2\font plus
\BIBentryALTinterwordstretchfactor\fontdimen3\font minus
  \fontdimen4\font\relax}
\providecommand{\BIBforeignlanguage}[2]{{%
\expandafter\ifx\csname l@#1\endcsname\relax
\typeout{** WARNING: IEEEtran.bst: No hyphenation pattern has been}%
\typeout{** loaded for the language `#1'. Using the pattern for}%
\typeout{** the default language instead.}%
\else
\language=\csname l@#1\endcsname
\fi
#2}}
\providecommand{\BIBdecl}{\relax}
\BIBdecl

\bibitem{w1}
\BIBentryALTinterwordspacing
``Cisco annual internet report white paper.'' [Online]. Available:
  \url{https://www.cisco.com/c/en/us/solutions/collateral/executive-perspectives/annual-internet-report/white-paper-c11-741490.html}
\BIBentrySTDinterwordspacing

\bibitem{smith2013live}
T.~Smith, M.~Obrist, and P.~Wright, ``Live-streaming changes the (video)
  game,'' in \emph{Proceedings of the 11th european conference on Interactive
  TV and video}, 2013, pp. 131--138.

\bibitem{avc}
D.~Marpe, T.~Wiegand, and G.~J. Sullivan, ``The h. 264/mpeg4 advanced video
  coding standard and its applications,'' \emph{IEEE communications magazine},
  vol.~44, no.~8, pp. 134--143, 2006.

\bibitem{hevc}
G.~J. Sullivan, J.-R. Ohm, W.-J. Han, and T.~Wiegand, ``Overview of the high
  efficiency video coding (hevc) standard,'' \emph{IEEE Transactions on
  circuits and systems for video technology}, vol.~22, no.~12, pp. 1649--1668,
  2012.

\bibitem{vvc}
B.~Bross, Y.-K. Wang, Y.~Ye, S.~Liu, J.~Chen, G.~J. Sullivan, and J.-R. Ohm,
  ``Overview of the versatile video coding (vvc) standard and its
  applications,'' \emph{IEEE Transactions on Circuits and Systems for Video
  Technology}, vol.~31, no.~10, pp. 3736--3764, 2021.

\bibitem{berger2003rate}
T.~Berger, ``Rate-distortion theory,'' \emph{Wiley Encyclopedia of
  Telecommunications}, 2003.

\bibitem{tang2019fast}
N.~Tang, J.~Cao, F.~Liang, J.~Wang, H.~Liu, X.~Wang, and X.~Du, ``Fast ctu
  partition decision algorithm for vvc intra and inter coding,'' in \emph{2019
  IEEE Asia Pacific Conference on Circuits and Systems (APCCAS)}.\hskip 1em
  plus 0.5em minus 0.4em\relax IEEE, 2019, pp. 361--364.

\bibitem{huang2011predictive}
Y.~Huang and R.~P. Rao, ``Predictive coding,'' \emph{Wiley Interdisciplinary
  Reviews: Cognitive Science}, vol.~2, no.~5, pp. 580--593, 2011.

\bibitem{malvar2003low}
H.~S. Malvar, A.~Hallapuro, M.~Karczewicz, and L.~Kerofsky, ``Low-complexity
  transform and quantization in h. 264/avc,'' \emph{IEEE Transactions on
  circuits and systems for video technology}, vol.~13, no.~7, pp. 598--603,
  2003.

\bibitem{budagavi2014hevc}
M.~Budagavi, A.~Fuldseth, and G.~Bj{\o}ntegaard, ``Hevc transform and
  quantization,'' in \emph{High Efficiency Video Coding (HEVC)}.\hskip 1em plus
  0.5em minus 0.4em\relax Springer, 2014, pp. 141--169.

\bibitem{norkin2012hevc}
A.~Norkin, G.~Bjontegaard, A.~Fuldseth, M.~Narroschke, M.~Ikeda, K.~Andersson,
  M.~Zhou, and G.~Van~der Auwera, ``Hevc deblocking filter,'' \emph{IEEE
  Transactions on Circuits and Systems for Video Technology}, vol.~22, no.~12,
  pp. 1746--1754, 2012.

\bibitem{sze2012high}
V.~Sze and M.~Budagavi, ``High throughput cabac entropy coding in hevc,''
  \emph{IEEE Transactions on Circuits and Systems for Video Technology},
  vol.~22, no.~12, pp. 1778--1791, 2012.

\bibitem{tm5}
L.~Wang, ``Rate control for mpeg video coding,'' \emph{Signal Processing: Image
  Communication}, vol.~15, no.~6, pp. 493--511, 2000.

\bibitem{tmn8}
J.-C. Tsai and C.-H. Shieh, ``Modified tmn8 rate control for low-delay video
  communications,'' \emph{IEEE transactions on circuits and systems for video
  technology}, vol.~14, no.~6, pp. 864--868, 2004.

\bibitem{jvt-g012}
Z.~Li and F.~Pan, ``Adaptive basic unit layer rate control for jvt
  (jvt-g012),'' in \emph{Joint Video Team (JVT) 7th Meeting at Pattaya}, March.
  2003.

\bibitem{y71}
H.~Choi, J.~Nam, J.~Yoo, and D.~Sim, ``Rate control based on unified rq model
  for hevc (jctvc-h0213),'' in \emph{Joint Collaborative Team on Video Coding
  (JCT-VC) 8th Meeting, San José, CA}, Feb. 2012.

\bibitem{k0103}
B.~Li, H.~Li, and L.~Li, ``Rate control by r-lambda model for hevc,'' in
  \emph{ITU-T/ISO/IEC JCT-VC Document JCTVC-K0103}, Oct. 2012.

\bibitem{y47}
Y.~Li and Z.~Chen, ``Rate control for vvc (jvet-k0390),'' in \emph{Joint Video
  Experts Team (JVET) 11th Meeting, Ljubljana, SI}, July 2018.

\bibitem{c11}
Y.~Liu, Z.~G. Li, and Y.~C. Soh, ``A novel rate control scheme for low delay
  video communication of h.264/avc standard,'' \emph{IEEE Transactions on
  Circuits and Systems for Video Technology}, vol.~17, no.~1, pp. 68--78, 2007.

\bibitem{c14}
S.~Hu, H.~Wang, S.~Kwong, and T.~Zhao, ``Frame level rate control for h.264/avc
  with novel rate-quantization model,'' in \emph{2010 IEEE International
  Conference on Multimedia and Expo}, 2010, pp. 226--231.

\bibitem{i0426}
B.~Li and D.~Zhang, ``Qp determination by lambda value,'' in \emph{Joint
  Collaborative Team on Video Coding (JCT-VC) 9th Meetting at Geneva}, Apr.
  2012.

\bibitem{y12}
M.~Karczewicz and X.~Wang, ``Intra frame rate control based on satd
  (jctvc-m0257),'' in \emph{Joint Collaborative Team on Video Coding (JCT-VC)
  13th Meeting, Incheon, KR}, Apr. 2013.

\bibitem{y14}
M.~Fang, M.~Tang, and J.~Wen, ``R-lambda model based rate control with
  pre-encoding process (jctvc-u0152),'' in \emph{Joint Collaborative Team on
  Video Coding (JCT-VC) 21st Meeting, Warsaw, PL}, June 2015.

\bibitem{w3}
Z.~He, Y.~K. Kim, and S.~Mitra, ``Low-delay rate control for dct video coding
  via /spl rho/-domain source modeling,'' \emph{IEEE Transactions on Circuits
  and Systems for Video Technology}, vol.~11, no.~8, pp. 928--940, 2001.

\bibitem{w4}
M.~Liu, Y.~Guo, H.~Li, and C.~W. Chen, ``Low-complexity rate control based on
  rho-domain model for scalable video coding,'' in \emph{2010 IEEE
  International Conference on Image Processing}, 2010, pp. 1277--1280.

\bibitem{c3}
H.-J. Lee, T.~Chiang, and Y.-Q. Zhang, ``Scalable rate control for mpeg-4
  video,'' \emph{IEEE Transactions on Circuits and Systems for Video
  Technology}, vol.~10, no.~6, pp. 878--894, 2000.

\bibitem{c4}
Z.~He and T.~Chen, ``Linear rate control for jvt video coding,'' in
  \emph{International Conference on Information Technology: Research and
  Education, 2003. Proceedings. ITRE2003.}, 2003, pp. 65--68.

\bibitem{c5}
M.~Jiang, X.~Yi, and N.~Ling, ``Improved frame-layer rate control for h.264
  using mad ratio,'' in \emph{2004 IEEE International Symposium on Circuits and
  Systems (ISCAS)}, vol.~3, 2004, pp. III--813.

\bibitem{c7}
J.~Xu and Y.~He, ``A novel rate control for h.264,'' in \emph{2004 IEEE
  International Symposium on Circuits and Systems (ISCAS)}, vol.~3, 2004, pp.
  III--809.

\bibitem{c8}
S.~Ma, W.~Gao, and Y.~Lu, ``Rate-distortion analysis for h.264/avc video coding
  and its application to rate control,'' \emph{IEEE Transactions on Circuits
  and Systems for Video Technology}, vol.~15, no.~12, pp. 1533--1544, 2005.

\bibitem{c9}
W.~Yuan, S.~Lin, Y.~Zhang, W.~Yuan, and H.~Luo, ``Optimum bit allocation and
  rate control for h.264/avc,'' \emph{IEEE Transactions on Circuits and Systems
  for Video Technology}, vol.~16, no.~6, pp. 705--715, 2006.

\bibitem{c10}
D.-K. Kwon, M.-Y. Shen, and C.-C.~J. Kuo, ``Rate control for h.264 video with
  enhanced rate and distortion models,'' \emph{IEEE Transactions on Circuits
  and Systems for Video Technology}, vol.~17, no.~5, pp. 517--529, 2007.

\bibitem{c12}
H.~Wang and S.~Kwong, ``Rate-distortion optimization of rate control for h.264
  with adaptive initial quantization parameter determination,'' \emph{IEEE
  Transactions on Circuits and Systems for Video Technology}, vol.~18, no.~1,
  pp. 140--144, 2008.

\bibitem{c13}
W.-J. Tsai and T.-L. Chou, ``Scene change aware intra-frame rate control for
  h.264/avc,'' \emph{IEEE Transactions on Circuits and Systems for Video
  Technology}, vol.~20, no.~12, pp. 1882--1886, 2010.

\bibitem{c15}
H.-M. Hu, B.~Li, W.~Lin, W.~Li, and M.-T. Sun, ``Region-based rate control for
  h.264/avc for low bit-rate applications,'' \emph{IEEE Transactions on
  Circuits and Systems for Video Technology}, vol.~22, no.~11, pp. 1564--1576,
  2012.

\bibitem{y22}
X.~Liang, Q.~Wang, Y.~Zhou, B.~Luo, and A.~Men, ``A novel rq model based rate
  control scheme in hevc,'' in \emph{2013 Visual Communications and Image
  Processing (VCIP)}.\hskip 1em plus 0.5em minus 0.4em\relax IEEE, 2013, pp.
  1--6.

\bibitem{y25}
S.~Wang, S.~Ma, L.~Zhang, S.~Wang, D.~Zhao, and W.~Gao, ``Multi layer based
  rate control algorithm for hevc,'' in \emph{2013 IEEE International Symposium
  on Circuits and Systems (ISCAS)}.\hskip 1em plus 0.5em minus 0.4em\relax
  IEEE, 2013, pp. 41--44.

\bibitem{y24}
L.~Tian, Y.~Zhou, and X.~Cao, ``A new rate-complexity-qp algorithm (rcqa) for
  hevc intra-picture rate control,'' in \emph{2014 International Conference on
  Computing, Networking and Communications (ICNC)}.\hskip 1em plus 0.5em minus
  0.4em\relax IEEE, 2014, pp. 375--380.

\bibitem{y27}
W.~Wu, J.~Liu, and L.~Feng, ``Novel rate control scheme for low delay video
  coding of hevc,'' \emph{ETRI Journal}, vol.~38, no.~1, pp. 185--194, 2016.

\bibitem{y23}
B.~Hosking, D.~Agrafiotis, D.~Bull, and N.~Eastern, ``An adaptive resolution
  rate control method for intra coding in hevc,'' in \emph{2016 IEEE
  International Conference on Acoustics, Speech and Signal Processing
  (ICASSP)}, 2016, pp. 1486--1490.

\bibitem{y56}
Y.~Mao, M.~Wang, S.~Wang, and S.~Kwong, ``High efficiency rate control for
  versatile video coding based on composite cauchy distribution,'' \emph{IEEE
  Transactions on Circuits and Systems for Video Technology}, vol.~32, no.~4,
  pp. 2371--2384, 2022.

\bibitem{y61}
C.~R. Helmrich, I.~Zupancic, J.~Brandenburg, V.~George, A.~Wieckowski, and
  B.~Bross, ``Visually optimized two-pass rate control for video coding using
  the low-complexity xpsnr model,'' in \emph{2021 International Conference on
  Visual Communications and Image Processing (VCIP)}.\hskip 1em plus 0.5em
  minus 0.4em\relax IEEE, 2021, pp. 1--5.

\bibitem{c1}
M.~Wang and B.~Yan, ``Lagrangian multiplier based joint three-layer rate
  control for h.264/avc,'' \emph{IEEE Signal Processing Letters}, vol.~16,
  no.~8, pp. 679--682, 2009.

\bibitem{y38}
B.~Lee, M.~Kim, and T.~Q. Nguyen, ``A frame-level rate control scheme based on
  texture and nontexture rate models for high efficiency video coding,''
  \emph{IEEE Transactions on Circuits and Systems for Video Technology},
  vol.~24, no.~3, pp. 465--479, 2013.

\bibitem{y1}
B.~Li, H.~Li, L.~Li, and J.~Zhang, ``$\lambda $ domain rate control algorithm
  for high efficiency video coding,'' \emph{IEEE Transactions on Image
  Processing}, vol.~23, no.~9, pp. 3841--3854, 2014.

\bibitem{y19}
M.~Meddeb, M.~Cagnazzo, and B.~Pesquet-Popescu, ``Region-of-interest-based rate
  control scheme for high-efficiency video coding,'' \emph{APSIPA Transactions
  on Signal and Information Processing}, vol.~3, 2014.

\bibitem{y44}
Z.~Yang, L.~Song, Z.~Luo, and X.~Wang, ``Low delay rate control for hevc,'' in
  \emph{2014 IEEE International Symposium on Broadband Multimedia Systems and
  Broadcasting}.\hskip 1em plus 0.5em minus 0.4em\relax IEEE, 2014, pp. 1--5.

\bibitem{y46}
M.~Zhou, X.~Wei, S.~Wang, S.~Kwong, C.-K. Fong, P.~H. Wong, W.~Y. Yuen, and
  W.~Gao, ``Ssim-based global optimization for ctu-level rate control in
  hevc,'' \emph{IEEE Transactions on Multimedia}, vol.~21, no.~8, pp.
  1921--1933, 2019.

\bibitem{y3}
M.~Wang, K.~N. Ngan, and H.~Li, ``An efficient frame-content based intra frame
  rate control for high efficiency video coding,'' \emph{IEEE Signal Processing
  Letters}, vol.~22, no.~7, pp. 896--900, 2015.

\bibitem{y20}
H.~Zeng, A.~Yang, K.~N. Ngan, and M.~Wang, ``Perceptual sensitivity-based rate
  control method for high efficiency video coding,'' \emph{Multimedia tools and
  applications}, vol.~75, no.~17, pp. 10\,383--10\,396, 2016.

\bibitem{y2}
M.~Zhou, Y.~Zhang, B.~Li, and H.-M. Hu, ``Complexity-based intra frame rate
  control by jointing inter-frame correlation for high efficiency video
  coding,'' \emph{Journal of Visual Communication and Image Representation},
  vol.~42, pp. 46--64, 2017.

\bibitem{y5}
S.~Li, M.~Xu, Z.~Wang, and X.~Sun, ``Optimal bit allocation for ctu level rate
  control in hevc,'' \emph{IEEE Transactions on Circuits and Systems for Video
  Technology}, vol.~27, no.~11, pp. 2409--2424, 2017.

\bibitem{y40}
M.~Wang, K.~N. Ngan, and H.~Li, ``Low-delay rate control for consistent quality
  using distortion-based lagrange multiplier,'' \emph{IEEE Transactions on
  Image Processing}, vol.~25, no.~7, pp. 2943--2955, 2016.

\bibitem{y15}
V.~Sanchez, ``Rate control for hevc intra-coding based on piecewise linear
  approximations,'' in \emph{2018 IEEE International Conference on Acoustics,
  Speech and Signal Processing (ICASSP)}, 2018, pp. 1782--1786.

\bibitem{y4}
Y.~Gong, S.~Wan, K.~Yang, H.~R. Wu, and Y.~Liu, ``Temporal-layer-motivated
  lambda domain picture level rate control for random-access configuration in
  h.265/hevc,'' \emph{IEEE Transactions on Circuits and Systems for Video
  Technology}, vol.~29, no.~1, pp. 156--170, 2019.

\bibitem{y17}
K.~R. Perez-Daniel and V.~Sanchez, ``Luma-aware multi-model rate-control for
  hdr content in hevc,'' in \emph{2017 IEEE International Conference on Image
  Processing (ICIP)}, 2017, pp. 1022--1026.

\bibitem{y37}
H.~Guo, C.~Zhu, S.~Li, and Y.~Gao, ``Optimal bit allocation at frame level for
  rate control in hevc,'' \emph{IEEE Transactions on Broadcasting}, vol.~65,
  no.~2, pp. 270--281, 2018.

\bibitem{y21}
W.~Li, P.~Ren, E.~Zhang, and F.~Zhao, ``Rate control for hevc intra-coding with
  a ctu-dependent distortion model,'' \emph{Signal, Image and Video
  Processing}, vol.~13, no.~1, pp. 17--25, 2019.

\bibitem{y16}
J.~Mir, D.~S. Talagala, and A.~Fernando, ``Optimization of hevc
  $\lambda$-domain rate control algorithm for hdr video,'' in \emph{2018 IEEE
  International Conference on Consumer Electronics (ICCE)}, 2018, pp. 1--4.

\bibitem{y39}
M.~Zhou, X.~Wei, S.~Wang, S.~Kwong, C.-K. Fong, P.~H. Wong, and W.~Y. Yuen,
  ``Global rate-distortion optimization-based rate control for hevc hdr
  coding,'' \emph{IEEE Transactions on Circuits and Systems for Video
  Technology}, vol.~30, no.~12, pp. 4648--4662, 2019.

\bibitem{y36}
W.~Lim and D.~Sim, ``A perceptual rate control algorithm based on luminance
  adaptation for hevc encoders,'' \emph{Signal, Image and Video Processing},
  vol.~14, no.~5, pp. 887--895, 2020.

\bibitem{y41}
Z.~Chen and X.~Pan, ``An optimized rate control for low-delay h. 265/hevc,''
  \emph{IEEE Transactions on Image Processing}, vol.~28, no.~9, pp. 4541--4552,
  2019.

\bibitem{y42}
M.~Zhou, X.~Wei, S.~Kwong, W.~Jia, and B.~Fang, ``Just noticeable
  distortion-based perceptual rate control in hevc,'' \emph{IEEE Transactions
  on Image Processing}, vol.~29, pp. 7603--7614, 2020.

\bibitem{y53}
M.~H. Hyun, B.~Lee, and M.~Kim, ``A frame-level constant bit-rate control using
  recursive bayesian estimation for versatile video coding,'' \emph{IEEE
  Access}, vol.~8, pp. 227\,255--227\,269, 2020.

\bibitem{y57}
Y.~Chen, S.~Kwong, M.~Zhou, S.~Wang, G.~Zhu, and Y.~Wang, ``Intra frame rate
  control for versatile video coding with quadratic rate-distortion
  modelling,'' in \emph{ICASSP 2020-2020 IEEE International Conference on
  Acoustics, Speech and Signal Processing (ICASSP)}.\hskip 1em plus 0.5em minus
  0.4em\relax IEEE, 2020, pp. 4422--4426.

\bibitem{y60}
Y.~Li, Z.~Liu, Z.~Chen, and S.~Liu, ``Rate control for versatile video
  coding,'' in \emph{2020 IEEE International Conference on Image Processing
  (ICIP)}.\hskip 1em plus 0.5em minus 0.4em\relax IEEE, 2020, pp. 1176--1180.

\bibitem{y59}
F.~Liu and Z.~Chen, ``Multi-objective optimization of quality in vvc rate
  control for low-delay video coding,'' \emph{IEEE Transactions on Image
  Processing}, vol.~30, pp. 4706--4718, 2021.

\bibitem{c2}
Z.~Li, F.~Pan, K.~Lim, X.~Lin, and S.~Rahardja, ``Adaptive rate control for
  h.264,'' in \emph{2004 International Conference on Image Processing, 2004.
  ICIP '04.}, vol.~2, 2004, pp. 745--748 Vol.2.

\bibitem{y33}
J.~Si, S.~Ma, X.~Zhang, and W.~Gao, ``Adaptive rate control for high efficiency
  video coding,'' in \emph{2012 Visual Communications and Image Processing},
  2012, pp. 1--6.

\bibitem{y43}
Y.-J. Yoon, H.~Kim, S.-h. Jung, D.~Jun, Y.~Kim, J.~S. Choi, and S.-J. Ko, ``A
  new rate control method for hierarchical video coding in hevc,'' in
  \emph{IEEE International symposium on Broadband Multimedia Systems and
  Broadcasting}.\hskip 1em plus 0.5em minus 0.4em\relax IEEE, 2012, pp. 1--4.

\bibitem{y35}
J.~Si, S.~Ma, and W.~Gao, ``Adaptive rate control for hevc (jctvc-j0057),'' in
  \emph{Joint Collaborative Team on Video Coding (JCT-VC) 10th Meeting,
  Stockholm, SE}, July 2012.

\bibitem{y6}
H.~Zhihai, Y.~Kim, and S.~Mitra, ``Low-delay rate control for dct video coding
  via p-domain source modeling,'' \emph{IEEE Trans on CSTV}, vol.~11, no.~8,
  pp. 813--816, 2001.

\bibitem{y7}
Z.~He, Y.~K. Kim, and S.~Mitra, ``$\rho$-domain source modeling and rate
  control for video coding and transmission,'' in \emph{2001 IEEE International
  Conference on Acoustics, Speech, and Signal Processing. Proceedings (Cat.
  No.01CH37221)}, vol.~3, 2001, pp. 1773--1776 vol.3.

\bibitem{c6}
I.-H. Shin, Y.-L. Lee, and H.~Park, ``Rate control using linear rate-$\rho$
  model for h.264,'' \emph{Signal Processing: Image Communication}, vol.~19,
  no.~4, pp. 341--352, 2004.

\bibitem{y8}
S.~Wang, S.~Ma, S.~Wang, D.~Zhao, and W.~Gao, ``Rate-gop based rate control for
  high efficiency video coding,'' \emph{IEEE Journal of Selected Topics in
  Signal Processing}, vol.~7, no.~6, pp. 1101--1111, 2013.

\bibitem{y28}
------, ``Quadratic $\rho$-domain based rate control algorithm for hevc,''
  \emph{2013 IEEE International Conference on Acoustics, Speech and Signal
  Processing}, pp. 1695--1699, 2013.

\bibitem{y26}
T.~Biatek, M.~Raulet, J.-F. Travers, and O.~Deforges, ``Efficient quantization
  parameter estimation in hevc based on $\rho$-domain,'' in \emph{European
  Signal Processing Conference}, 2014.

\bibitem{y11}
A.~A. Ramanand, I.~Ahmad, and V.~Swaminathan, ``A survey of rate control in
  hevc and shvc video encoding,'' in \emph{2017 IEEE International Conference
  on Multimedia \& Expo Workshops (ICMEW)}, 2017, pp. 145--150.

\bibitem{y13}
J.~Wen, M.~Fang, and M.~Tang, ``R-lambda model based rate control with
  pre-encoding process (jctvc-t0216),'' in \emph{Joint Collaborative Team on
  Video Coding (JCT-VC) 20th Meeting, vGeneva, CH}, Feb. 2015.

\bibitem{rminorr204}
\BIBentryALTinterwordspacing
M.~Tang, J.~Wen, and Y.~Han, ``A generalized rate-distortion-{\(\lambda\)}
  model based {HEVC} rate control algorithm,'' \emph{CoRR}, vol.
  abs/1911.00639, 2019. [Online]. Available:
  \url{http://arxiv.org/abs/1911.00639}
\BIBentrySTDinterwordspacing

\bibitem{rminorr201}
T.~Li, L.~Yu, H.~Wang, and Z.~Kuang, ``A bit allocation method based on
  inter-view dependency and spatio-temporal correlation for multi-view texture
  video coding,'' \emph{IEEE Transactions on Broadcasting}, vol.~67, no.~1, pp.
  159--173, 2021.

\bibitem{rminorr203}
H.~Guo, C.~Zhu, Y.~Gao, and S.~Song, ``A frame-level rate control scheme for
  low delay video coding in hevc,'' in \emph{2017 IEEE 19th International
  Workshop on Multimedia Signal Processing (MMSP)}, 2017, pp. 1--6.

\bibitem{y50}
Z.~Liu, Z.~Chen, and Y.~Li, ``Ahg10: Quality dependency factor based rate
  control for vvc (jvet-m0600),'' in \emph{Joint Video Experts Team (JVET) 13th
  Meeting, Marrakech, MA}, Jan. 2019.

\bibitem{y51}
F.~Liu, Z.~Liu, Y.~Li, and Z.~Chen, ``Ahg10: Extension of rate control to
  support random access configuration with gop size of 32 (jvet-t0062),'' in
  \emph{Joint Video Experts Team (JVET) 20th Meeting by teleconference}, Oct.
  2020.

\bibitem{y52}
G.~Ren, J.~Jia, J.~Wang, and Z.~Chen, ``Ahg10: An improved vvc rate control
  scheme ( jvet-y0105),'' in \emph{Joint Video Experts Team (JVET) 21th
  Meeting, by teleconference}, Jan. 2022.

\bibitem{y30}
W.~Gao, S.~Kwong, and Y.~Jia, ``Joint machine learning and game theory for rate
  control in high efficiency video coding,'' \emph{IEEE Transactions on Image
  Processing}, vol.~26, no.~12, pp. 6074--6089, 2017.

\bibitem{y29}
J.-H. Hu, W.-H. Peng, and C.-H. Chung, ``Reinforcement learning for hevc/h. 265
  intra-frame rate control,'' in \emph{2018 IEEE International Symposium on
  Circuits and Systems (ISCAS)}.\hskip 1em plus 0.5em minus 0.4em\relax IEEE,
  2018, pp. 1--5.

\bibitem{y31}
M.~Zhou, X.~Wei, S.~Kwong, W.~Jia, and B.~Fang, ``Rate control method based on
  deep reinforcement learning for dynamic video sequences in hevc,'' \emph{IEEE
  Transactions on Multimedia}, vol.~23, pp. 1106--1121, 2020.

\bibitem{y32}
X.~Wei, M.~Zhou, S.~Kwong, H.~Yuan, and T.~Xiang, ``Joint reinforcement
  learning and game theory bitrate control method for 360-degree dynamic
  adaptive streaming,'' in \emph{ICASSP 2021-2021 IEEE International Conference
  on Acoustics, Speech and Signal Processing (ICASSP)}.\hskip 1em plus 0.5em
  minus 0.4em\relax IEEE, 2021, pp. 4230--4234.

\bibitem{y54}
F.~Raufmehr, M.~R. Salehi, and E.~Abiri, ``A frame-level mlp-based bit-rate
  controller for real-time video transmission using vvc standard,''
  \emph{Journal of Real-Time Image Processing}, vol.~18, no.~3, pp. 751--763,
  2021.

\bibitem{y55}
R.~Farhad, S.~M. Reza, and A.~Ebrahim, ``A neural network-based video bit-rate
  control algorithm for variable bit-rate applications of versatile video
  coding standard,'' \emph{Signal Processing: Image Communication}, vol.~96, p.
  116317, 2021.

\bibitem{y58}
M.~Wang, J.~Zhang, L.~Huang, and J.~Xiong, ``Machine learning-based rate
  distortion modeling for vvc/h. 266 intra-frame,'' in \emph{2021 IEEE
  International Conference on Multimedia and Expo (ICME)}.\hskip 1em plus 0.5em
  minus 0.4em\relax IEEE, 2021, pp. 1--6.

\bibitem{c16}
L.~Xu, S.~Ma, D.~Zhao, and W.~Gao, ``{Rate control for scalable video model},''
  in \emph{Visual Communications and Image Processing 2005}, S.~Li, F.~Pereira,
  H.-Y. Shum, and A.~G. Tescher, Eds., vol. 5960, International Society for
  Optics and Photonics.\hskip 1em plus 0.5em minus 0.4em\relax SPIE, 2005, pp.
  525 -- 534.

\bibitem{c17}
Y.~Liu, Z.~G. Li, and Y.~C. Soh, ``Rate control of h.264/avc scalable
  extension,'' \emph{IEEE Transactions on Circuits and Systems for Video
  Technology}, vol.~18, no.~1, pp. 116--121, 2008.

\bibitem{c18}
Y.~Pitrey, M.~Babel, O.~Déforges, and J.~Viéron, ``{$\rho$-domain based rate
  control scheme for spatial, temporal, and quality scalable video coding},''
  in \emph{Visual Communications and Image Processing 2009}, M.~Rabbani and
  R.~L. Stevenson, Eds., vol. 7257, International Society for Optics and
  Photonics.\hskip 1em plus 0.5em minus 0.4em\relax SPIE, 2009, pp. 25 -- 32.

\bibitem{y34}
S.~Hu, H.~Wang, S.~Kwong, T.~Zhao, and C.-C.~J. Kuo, ``Rate control
  optimization for temporal-layer scalable video coding,'' \emph{IEEE
  Transactions on Circuits and Systems for Video Technology}, vol.~21, no.~8,
  pp. 1152--1162, 2011.

\bibitem{y49}
Y.~Li, D.~Liu, and Z.~Chen, ``Ahg9-related: Cnn-based lambda-domain rate
  control for intra frames (jvet-m0215),'' in \emph{Joint Video Experts Team
  (JVET) 13th Meeting, Marrakech, MA}, Jan. 2019.

\bibitem{y64}
G.~Lu, W.~Ouyang, D.~Xu, X.~Zhang, C.~Cai, and Z.~Gao, ``Dvc: An end-to-end
  deep video compression framework,'' in \emph{Proceedings of the IEEE/CVF
  Conference on Computer Vision and Pattern Recognition}, 2019, pp.
  11\,006--11\,015.

\bibitem{c19}
X.~Jing, L.-P. Chau, and W.-C. Siu, ``Frame complexity-based rate-quantization
  model for h.264/avc intraframe rate control,'' \emph{IEEE Signal Processing
  Letters}, vol.~15, pp. 373--376, 2008.

\bibitem{c20}
Z.~Liu, H.~Yan, L.~Shen, Y.~Wang, and Z.~Zhang, ``A motion attention model
  based rate control algorithm for h.264/avc,'' in \emph{2009 Eighth IEEE/ACIS
  International Conference on Computer and Information Science}, 2009, pp.
  568--573.

\bibitem{c21}
L.~Shen, Z.~Liu, and Z.~Zhang, ``A novel h.264 rate control algorithm with
  consideration of visual attention,'' \emph{Multimed Tools Appl}, vol.~63, p.
  709–727, 2013.

\bibitem{y10}
J.~Ribas-Corbera and S.~Lei, ``Rate control in dct video coding for low-delay
  communications,'' \emph{IEEE Transactions on Circuits and Systems for Video
  Technology}, vol.~9, no.~1, pp. 172--185, 1999.

\bibitem{y45}
C.-W. Seo, J.-H. Moon, and J.-K. Han, ``Rate control for consistent objective
  quality in high efficiency video coding,'' \emph{IEEE transactions on image
  processing}, vol.~22, no.~6, pp. 2442--2454, 2013.

\bibitem{wlie2005two}
W.-N. Lie, C.-F. Chen, and T.~C.-I. Lin, ``Two-pass rate-distortion optimized
  rate control technique for h. 264/avc video,'' in \emph{Visual Communications
  and Image Processing 2005}, vol. 5960.\hskip 1em plus 0.5em minus 0.4em\relax
  International Society for Optics and Photonics, 2005, p. 596035.

\bibitem{7340805}
S.~Wang, A.~Rehman, K.~Zeng, and Z.~Wang, ``Ssim-inspired two-pass rate control
  for high efficiency video coding,'' in \emph{2015 IEEE 17th International
  Workshop on Multimedia Signal Processing (MMSP)}, 2015, pp. 1--5.

\bibitem{rminorr202}
I.~Zupancic, M.~Naccari, M.~Mrak, and E.~Izquierdo, ``Two-pass rate control for
  improved quality of experience in uhdtv delivery,'' \emph{IEEE Journal of
  Selected Topics in Signal Processing}, vol.~11, no.~1, pp. 167--179, 2017.

\bibitem{meenderinck2009parallel}
C.~Meenderinck, A.~Azevedo, B.~Juurlink, M.~Alvarez~Mesa, and A.~Ramirez,
  ``Parallel scalability of video decoders,'' \emph{Journal of Signal
  Processing Systems}, vol.~57, no.~2, pp. 173--194, 2009.

\bibitem{ling2013efficiency}
Z.~Ling, X.~J. Jiang, and J.~J. Liu, ``Efficiency of dynamic gop length in
  video stream,'' in \emph{Advanced Materials Research}, vol. 765.\hskip 1em
  plus 0.5em minus 0.4em\relax Trans Tech Publ, 2013, pp. 879--884.

\bibitem{roitzsch2007slice}
M.~Roitzsch, ``Slice-balancing h. 264 video encoding for improved scalability
  of multicore decoding,'' in \emph{Proceedings of the 7th ACM \& IEEE
  international conference on Embedded software}, 2007, pp. 269--278.

\bibitem{blumenberg2013adaptive}
C.~Blumenberg, D.~Palomino, S.~Bampi, and B.~Zatt, ``Adaptive content-based
  tile partitioning algorithm for the hevc standard,'' in \emph{2013 Picture
  Coding Symposium (PCS)}.\hskip 1em plus 0.5em minus 0.4em\relax IEEE, 2013,
  pp. 185--188.

\bibitem{koziri2018combining}
M.~Koziri, P.~K. Papadopoulos, and T.~Loukopoulos, ``Combining tile parallelism
  with slice partitioning in video coding,'' in \emph{Applications of Digital
  Image Processing XLI}, vol. 10752.\hskip 1em plus 0.5em minus 0.4em\relax
  International Society for Optics and Photonics, 2018, p. 107520N.

\bibitem{karczewicz2021vvc}
M.~Karczewicz, N.~Hu, J.~Taquet, C.-Y. Chen, K.~Misra, K.~Andersson, P.~Yin,
  T.~Lu, E.~Fran{\c{c}}ois, and J.~Chen, ``Vvc in-loop filters,'' \emph{IEEE
  Transactions on Circuits and Systems for Video Technology}, vol.~31, no.~10,
  pp. 3907--3925, 2021.

\bibitem{van2003mapping}
E.~B. Van Der~Tol, E.~G. Jaspers, and R.~H. Gelderblom, ``Mapping of h. 264
  decoding on a multiprocessor architecture,'' in \emph{Image and Video
  Communications and Processing 2003}, vol. 5022.\hskip 1em plus 0.5em minus
  0.4em\relax International Society for Optics and Photonics, 2003, pp.
  707--718.

\bibitem{alvarez2012improving}
M.~Alvarez-Mesa, V.~George, T.~Schierl, and B.~Juurlink, ``Improving
  parallelization efficiency of wpp using overlapped wavefront,'' \emph{Joint
  Collaborative Team on Video Coding (JCT-VC), Document JCTVC-J0425,
  Stockholm}, vol.~3, 2012.

\bibitem{clare2012hevc}
G.~Clare and F.~Henry, ``An hevc transcoder converting non-parallel bitstreams
  to/from wpp,'' \emph{Joint Collaborative Team on Video Coding (JCT-VC),
  Document JCTVC-J0032, Stockholm}, 2012.

\bibitem{WWP01}
J.~Sainio, A.~Mercat, and J.~Vanne, ``Parallel implementations of lambda domain
  and r-lambda model rate control schemes in a practical hevc encoder,'' in
  \emph{2021 Data Compression Conference (DCC)}, 2021, pp. 368--368.

\bibitem{WWP02}
P.~Xu, K.~Chen, J.~Sun, X.~Ji, and Z.~Guo, ``An adaptive intra-frame parallel
  method based on complexity estimation for hevc,'' in \emph{2016 Visual
  Communications and Image Processing (VCIP)}, 2016, pp. 1--4.

\bibitem{WWP03}
S.~Wang, S.~Zhang, J.~Wang, L.~Chang, L.~Feng, and X.~Fan, ``Hardware
  architecture design of hevc entropy decoding,'' in \emph{2021 IEEE Intl Conf
  on Parallel \& Distributed Processing with Applications, Big Data \& Cloud
  Computing, Sustainable Computing \& Communications, Social Computing \&
  Networking (ISPA/BDCloud/SocialCom/SustainCom)}, 2021, pp. 1143--1150.

\bibitem{xu2015new}
S.~Xu, M.~Yu, S.~Fang, Z.~Peng, X.~Wang, and G.~Jiang, ``New rate control
  optimization algorithm for hevc aiming at discontinuous scene,'' \emph{WSEAS
  Transactions on computers}, vol.~14, no.~1, pp. 598--606, 2015.

\bibitem{song2017new}
F.~Song, C.~Zhu, Y.~Liu, and Y.~Zhou, ``A new gop level bit allocation method
  for hevc rate control,'' in \emph{2017 IEEE International Symposium on
  Broadband Multimedia Systems and Broadcasting (BMSB)}.\hskip 1em plus 0.5em
  minus 0.4em\relax IEEE, 2017, pp. 1--4.

\bibitem{li2016lambda}
L.~Li, B.~Li, H.~Li, and C.~W. Chen, ``$\lambda $ domain optimal bit allocation
  algorithm for high efficiency video coding,'' \emph{IEEE Transactions on
  Circuits and Systems for Video Technology}, vol.~28, no.~1, pp. 130--142,
  2016.

\bibitem{y66}
H.~Schwarz, D.~Marpe, and T.~Wiegand, ``Overview of the scalable video coding
  extension of the h.264/avc standard,'' \emph{IEEE Transactions on Circuits
  and Systems for Video Technology}, vol.~17, no.~9, pp. 1103--1120, 2007.

\bibitem{y69}
M.~Flierl and B.~Girod, ``Generalized b pictures and the draft h.264/avc
  video-compression standard,'' \emph{IEEE Transactions on Circuits and Systems
  for Video Technology}, vol.~13, no.~7, pp. 587--597, 2003.

\bibitem{hwang1993advanced}
K.~Hwang and N.~Jotwani, \emph{Advanced computer architecture: parallelism,
  scalability, programmability}.\hskip 1em plus 0.5em minus 0.4em\relax
  McGraw-Hill New York, 1993, vol. 199.

\bibitem{bryant2003computer}
R.~E. Bryant, O.~David~Richard, and O.~David~Richard, \emph{Computer systems: a
  programmer's perspective}.\hskip 1em plus 0.5em minus 0.4em\relax Prentice
  Hall Upper Saddle River, 2003, vol.~2.

\bibitem{ranger2007evaluating}
C.~Ranger, R.~Raghuraman, A.~Penmetsa, G.~Bradski, and C.~Kozyrakis,
  ``Evaluating mapreduce for multi-core and multiprocessor systems,'' in
  \emph{2007 IEEE 13th International Symposium on High Performance Computer
  Architecture}.\hskip 1em plus 0.5em minus 0.4em\relax Ieee, 2007, pp. 13--24.

\bibitem{CTC}
F.~Bossen, J.~Boyce, K.~Suehring, X.~Li, and V.~Seregin, ``Jvet common test
  conditions and software reference configurations for sdr video (joint video
  experts team (jvet) of itu-t sg 16 wp 3 and iso/iec jtc 1/sc 29/wg 11,
  jvet-n1010-v1),'' in \emph{Joint Video Exploration Team (JVET) 14th Meeting,
  Geneva, CH}, Apr. 2019.

\bibitem{jvt-h1100}
F.~Bossen, ``Common conditions and software reference configurations
  (jvt-j1100),'' in \emph{Joint Collaborative Team on Video Coding (JCT-VC) 8th
  Meeting}, Feb. 2012.

\bibitem{bjontegaard2001calculation}
G.~Bjontegaard, ``Calculation of average psnr differences between rd-curves,''
  \emph{VCEG-M33}, 2001.

\bibitem{c24}
L.~Bai, L.~Song, R.~Xie, L.~Zhang, and Z.~Luo, ``Rate control model for high
  dynamic range video,'' in \emph{2017 IEEE Visual Communications and Image
  Processing (VCIP)}, 2017, pp. 1--4.

\bibitem{c22}
J.~Kim, S.-H. Bae, and M.~Kim, ``An hevc-compliant perceptual video coding
  scheme based on jnd models for variable block-sized transform kernels,''
  \emph{IEEE Transactions on Circuits and Systems for Video Technology},
  vol.~25, no.~11, pp. 1786--1800, 2015.

\bibitem{c23}
H.~Wei, X.~Zhou, W.~Zhou, C.~Yan, Z.~Duan, and N.~Shan, ``Visual saliency based
  perceptual video coding in hevc,'' in \emph{2016 IEEE International Symposium
  on Circuits and Systems (ISCAS)}, 2016, pp. 2547--2550.

\bibitem{y68}
\BIBentryALTinterwordspacing
``Jm 9.4.'' [Online]. Available:
  \url{http://iphome.hhi.de/suehring/tml/download}
\BIBentrySTDinterwordspacing

\bibitem{rw10}
M.~Zhou, X.~Wei, C.~Ji, T.~Xiang, and B.~Fang, ``Optimum quality control
  algorithm for versatile video coding,'' \emph{IEEE Transactions on
  Broadcasting}, pp. 1--12, 2022.

\bibitem{rw02}
X.~Meng, C.~Jia, X.~Zhang, S.~Wang, and S.~Ma, ``Spatio-temporal correlation
  guided geometric partitioning for versatile video coding,'' \emph{IEEE
  Transactions on Image Processing}, vol.~31, pp. 30--42, 2022.

\bibitem{rw03}
Z.~Huang, K.~Lin, C.~Jia, S.~Wang, and S.~Ma, ``Beyond vvc: Towards perceptual
  quality optimized video compression using multi-scale hybrid approaches,'' in
  \emph{2021 IEEE/CVF Conference on Computer Vision and Pattern Recognition
  Workshops (CVPRW)}, 2021, pp. 1866--1869.

\bibitem{rw21b}
M.~Zhou, X.~Wei, W.~Jia, and S.~Kwong, ``Joint decision tree and visual feature
  rate control optimization for vvc uhd coding,'' \emph{IEEE Transactions on
  Image Processing}, vol.~32, pp. 219--234, 2023.

\bibitem{rw04}
B.~Bross, J.~Chen, J.-R. Ohm, G.~J. Sullivan, and Y.-K. Wang, ``Developments in
  international video coding standardization after avc, with an overview of
  versatile video coding (vvc),'' \emph{Proceedings of the IEEE}, vol. 109,
  no.~9, pp. 1463--1493, 2021.

\bibitem{rw06}
M.~Benjak, H.~Meuel, T.~Laude, and J.~Ostermann, ``Enhanced machine
  learning-based inter coding for vvc,'' in \emph{2021 International Conference
  on Artificial Intelligence in Information and Communication (ICAIIC)}, 2021,
  pp. 021--025.

\bibitem{rw07}
M.~Saldanha, G.~Sanchez, C.~Marcon, and L.~Agostini, ``Configurable fast block
  partitioning for vvc intra coding using light gradient boosting machine,''
  \emph{IEEE Transactions on Circuits and Systems for Video Technology}, pp.
  1--1, 2021.

\bibitem{rw08}
K.~Fischer, C.~Herglotz, and A.~Kaup, ``On versatile video coding at uhd with
  machine-learning-based super-resolution,'' in \emph{2020 Twelfth
  International Conference on Quality of Multimedia Experience (QoMEX)}, 2020,
  pp. 1--6.

\bibitem{rw09}
N.~Le, H.~Zhang, F.~Cricri, R.~Ghaznavi-Youvalari, H.~R. Tavakoli, and
  E.~Rahtu, ``Learned image coding for machines: A content-adaptive approach,''
  in \emph{2021 IEEE International Conference on Multimedia and Expo (ICME)},
  2021, pp. 1--6.

\bibitem{rw13}
H.~Wang, L.~Yu, J.~Liang, H.~Yin, T.~Li, and S.~Wang, ``Hierarchical predictive
  coding-based jnd estimation for image compression,'' \emph{IEEE Transactions
  on Image Processing}, vol.~30, pp. 487--500, 2021.

\bibitem{rw14}
Y.~Li and X.~Mou, ``Joint optimization for ssim-based ctu-level bit allocation
  and rate distortion optimization,'' \emph{IEEE Transactions on Broadcasting},
  vol.~67, no.~2, pp. 500--511, 2021.

\bibitem{rw15}
A.~Nakhaei and M.~Rezaei, ``Scene-level two-pass video rate controller for h.
  265/hevc standard,'' \emph{Multimedia Tools and Applications}, vol.~80,
  no.~5, pp. 7023--7038, 2021.

\bibitem{rw16}
J.~Chen, X.~Luo, M.~Hu, D.~Wu, and Y.~Zhou, ``Sparkle: User-aware viewport
  prediction in 360-degree video streaming,'' \emph{IEEE Transactions on
  Multimedia}, vol.~23, pp. 3853--3866, 2021.

\bibitem{rw17}
F.~Chiariotti, ``A survey on 360-degree video: Coding, quality of experience
  and streaming,'' \emph{Comput. Commun.}, vol. 177, pp. 133--155, 2021.

\bibitem{rw18}
T.~Zhao, J.~Lin, Y.~Song, X.~Wang, and Y.~Niu, \emph{Game Theory-Driven Rate
  Control for 360-Degree Video Coding}.\hskip 1em plus 0.5em minus 0.4em\relax
  New York, NY, USA: Association for Computing Machinery, 2021, p. 3998–4006.

\bibitem{rw19}
V.~Sanchez, ``Rate control for predictive transform screen content video coding
  based on ransac,'' \emph{IEEE Transactions on Circuits and Systems for Video
  Technology}, vol.~31, no.~11, pp. 4422--4438, 2021.

\bibitem{rw20}
K.~Perez-Daniel, F.~Garcia-Ugalde, and V.~Sanchez, ``Scene-based
  imperceptible-visible watermarking for hdr video content,'' in \emph{2019 7th
  International Workshop on Biometrics and Forensics (IWBF)}, 2019, pp. 1--6.

\bibitem{rw21}
H.~Yuan, Q.~Wang, Q.~Liu, J.~Huo, and P.~Li, ``Hybrid distortion-based
  rate-distortion optimization and rate control for h.265/hevc,'' \emph{IEEE
  Transactions on Consumer Electronics}, vol.~67, no.~2, pp. 97--106, 2021.

\bibitem{rminorr301}
Q.~Zhang, H.~Meng, Z.~Feng, and Z.~Han, ``Resource scheduling of time-sensitive
  services for b5g/6g connected automated vehicles,'' \emph{IEEE Internet of
  Things Journal}, pp. 1--1, 2022.

\bibitem{rminorr302}
Y.~Cui, Q.~Zhang, Z.~Feng, Z.~Wei, C.~Shi, J.~Fan, and P.~Zhang, ``Dual
  identities enabled low-latency visual networking for uav emergency
  communication,'' in \emph{GLOBECOM 2022 - 2022 IEEE Global Communications
  Conference}, 2022, pp. 474--479.

\bibitem{rminorr303}
A.~M. Girgis, J.~Park, M.~Bennis, and M.~Debbah, ``Predictive control and
  communication co-design via two-way gaussian process regression and aoi-aware
  scheduling,'' \emph{IEEE Transactions on Communications}, vol.~69, no.~10,
  pp. 7077--7093, 2021.

\end{thebibliography}
\end{document}